\documentclass[a4paper,11pt]{article}
\pdfoutput=1 % if your are submitting a pdflatex (i.e. if you have
\usepackage{tikz}
\usepackage{feynmp-auto}
\usepackage{tikz-feynman} % for Feynman diagrams
\usepackage{float}
\usepackage{jheppub} % for details on the use of the package, please
\usepackage{parskip}
\usepackage[T1]{fontenc} % if needed
\graphicspath{{img/}} % set images path

\title{\boldmath {Enhanced sensitivity to the $H \to Z\gamma \to \ell^+\ell^-\gamma$ decay at the LHC using machine learning and novel kinematic observables}}

\author[a]{Manisha Kumari\note{Corresponding author.}}
\author[a]{Amal Sarkar}

\affiliation[a]{Indian Institute of Technology Mandi, Kamand, Mandi, Himachal Pradesh 175005, India}

\emailAdd{manisha.kumari@cern.ch}

\abstract{
\noindent
At LHC energies, the Drell-Yan ($Z/\gamma^{*}$) processes have a substantially large cross section. Their di-lepton ($\ell^+\ell^-$) final state contributes significantly to many resonant signal regions, making them one of the dominant backgrounds in numerous physics analyses. The study focuses on improving the discrimination and suppression of the $Z/\gamma^{*} \rightarrow \ell^{+}\ell^{-}$ background from the $H \rightarrow Z\gamma \rightarrow \ell^{+}\ell^{-}\gamma$ signal at $\sqrt{s}=13~\text{TeV}$ by leveraging Monte Carlo simulated data. The analysis introduces physics-motivated correlated observables derived from the two-dimensional $(P_{\mathrm{Higgs}}, \theta_{Z\gamma})$ plane. These observables encode differences in angular and momentum information to enhance signal--background separation while maintaining high signal efficiency. We present a multivariate analysis (MVA) employing a Boosted Decision Tree (XGBoost) classifier. By incorporating additional physics-motivated correlated observables, the classifier achieves measurable improvements in performance. A significant increase in the area under the ROC curve (AUC) is observed in both the electron and muon channels, demonstrating the effectiveness of the expanded feature set. Further, optimised background rejection using $(P_{\mathrm{Higgs}}, \theta_{Z\gamma})$ plane increases the signal-to-background ratio to 2.1\% and 3.4\% for the electron and muon channels, respectively, near the Higgs mass. This work demonstrates that combining kinematic correlations with interpretable multivariate techniques improves sensitivity and robust background rejection. The approach is flexible and can be readily applied to a wide range of analyses, including rare Higgs decays, resonant searches, and studies beyond the Standard Model.}

\begin{document} 
\maketitle
\flushbottom

%======================
\section{Introduction}
\label{sec:intro}

\begin{figure}[hbt!]
  \centering
   \includegraphics[width=0.7\textwidth]{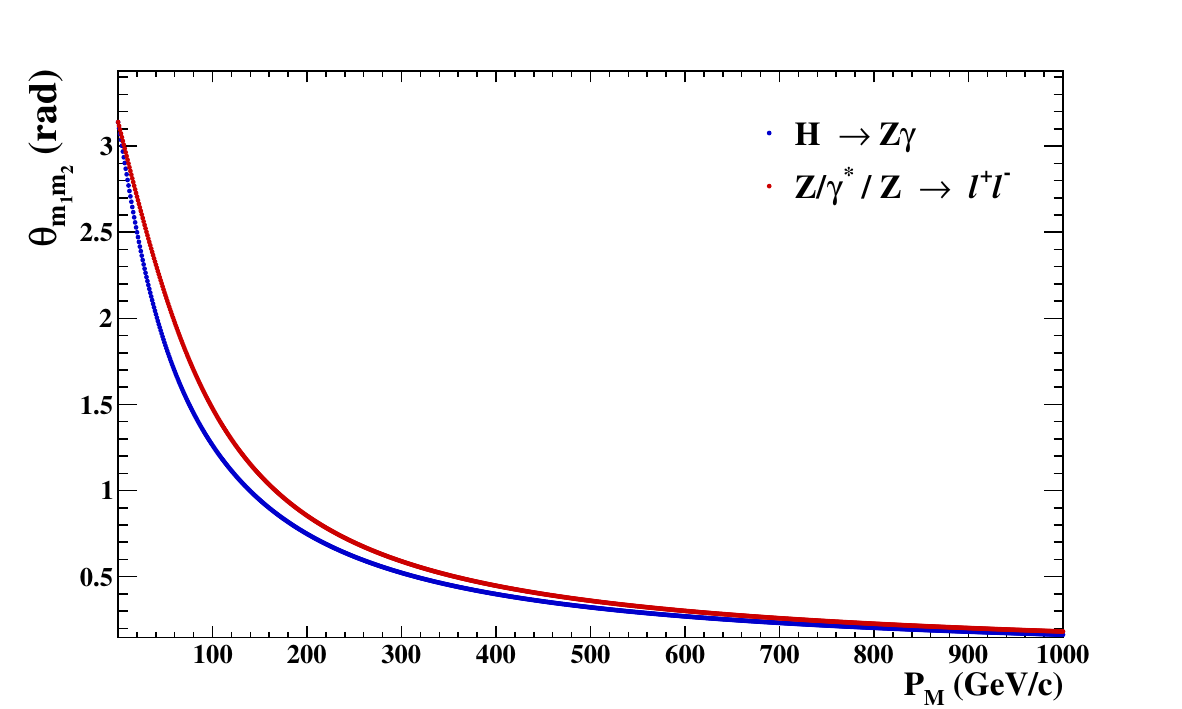}
\caption{(color online) Analytic two-body decay model illustrating the decrease of the opening angle $\theta_{\mathrm{open}}$ between the decay products as a function of the parent particle momentum $P_M$, arising from Lorentz boost effects. The curves correspond to decay configurations relevant for $H\to Z\gamma$ and $Z/\gamma^\ast \to \ell^+\ell^-$. Here, $m_1$ and $m_2$ denote the masses of the two final-state particles originating from the parent decay. For example, in $H\to Z\gamma$, $m_1 = m_Z$ and $m_2 = m_\gamma$, while for $Z/\gamma^\ast \to \ell^+\ell^-$, $m_1 = m_2 = m_\ell$. This description applies equally to both electron and muon final states.}
  \label{fig:theta_open_vs_PM_math}
\end{figure}

The Standard Model (SM) of particle physics provides a successful and consistent framework for describing the electroweak and strong interactions among fundamental particles~\cite{GLASHOW1961579, PhysRevLett.19.1264, Salam1968, Peskin1995, Donoghue1992,Pich:2012sx}. A key element of the SM is the mechanism of electroweak symmetry breaking realised through the Higgs field, which generates particle masses via the Higgs mechanism~\cite{Higgs:1964ia, Higgs1964, EWSB2012, Higgs1966}. The discovery of the Higgs boson by the ATLAS and CMS collaborations in 2012 at a mass of approximately 125~GeV completed the particle content of the SM~\cite{Aad2012, Chatrchyan2012}. The finding opened a new era of precision measurements of its properties and couplings.  Although the dominant Higgs boson production and decay modes along with its coupling with other particles have been extensively studied with high precision~\cite{hig_mes1, hig_mes2, hig_mes3, hig_mes4, hig_mes5, hig_mes6, hig_mes7, hig_mes8,hig_mes9}, rare decay channels remain only loosely constrained.
These channels provide complementary sensitivity to potential new-physics effects. Among these, the $H\rightarrow Z\gamma$ decay mode is highly sensitive to new physics effects as it proceeds exclusively through loop processes involving charged fermions and $W$ bosons, making it sensitive to potential contributions from extensions beyond the Standard Model~\cite{rad_higg_decay}. The predicted SM branching fraction for $H\rightarrow Z\gamma$ is approximately $1.5 \times 10^{-3}$~\cite{Branching_fraction}, significantly smaller than the dominant decay modes. An accurate measurement of this branching fraction provides a valuable probe of electroweak radiative corrections and potential deviations from SM predictions~\cite{rad_corr}. The $H \rightarrow Z\gamma \rightarrow \ell^{+}\ell^{-}\gamma$ final state is experimentally clean, consisting of two isolated same-flavour opposite-sign leptons and a high-energy photon~\cite{Hzgamma_atlas, Hzgamma_cms}. However, the primary experimental challenge in this channel arises from the overwhelming Drell-Yan background.
The $Z/\gamma^{*} \rightarrow \ell^{+}\ell^{-}$ process has a production cross section $\simeq 2~\mathrm{nb}$, which is significantly larger than that of the signal at the LHC energies~\cite{DrellYan1, DrellYan2_cms, DrellYan3_atlas}. In Drell-Yan events, photons arise from radiative processes or from misidentified objects, leading to reconstructed $\ell^{+}\ell^{-}\gamma$ final states that closely mimic the signal topology. As a consequence, the invariant mass spectrum of the $\ell^{+}\ell^{-}\gamma$ system observed in the data is dominated by a smooth, rapidly falling Drell-Yan background continuum, within which the Higgs boson signal would appear as a narrow resonance near 125~GeV. The large Drell-Yan yield and its kinematic overlap with the signal significantly limit the effectiveness of simple cut-based selections, motivating the need for more sophisticated analysis techniques. In this study, an approach is employed to develop physics-motivated correlated observables to enhance signal–background discrimination beyond the use of one-dimensional variables. The method is based on the fundamental kinematics of two-body decay: in the rest frame of a parent particle, the decay products are emitted back-to-back, while a Lorentz boost to the laboratory frame progressively collimates them, reducing the observed opening angle as the parent momentum increases. The degree of collimation depends on the mass scale of the decay products and resonance, leading to characteristic correlations between opening angles and reconstructed momenta that differ across distinct processes. To utilise these differences, an analytic decay calculation for various processes has been performed and presented in Figure~\ref{fig:theta_open_vs_PM_math}, which provides the baseline motivation for constructing new observables~\cite{twoBodyDecayNote}.

\begin{figure}[hbt]

    \centering
    \begin{minipage}{0.48\textwidth}
        \centering
        \includegraphics[width=\textwidth]{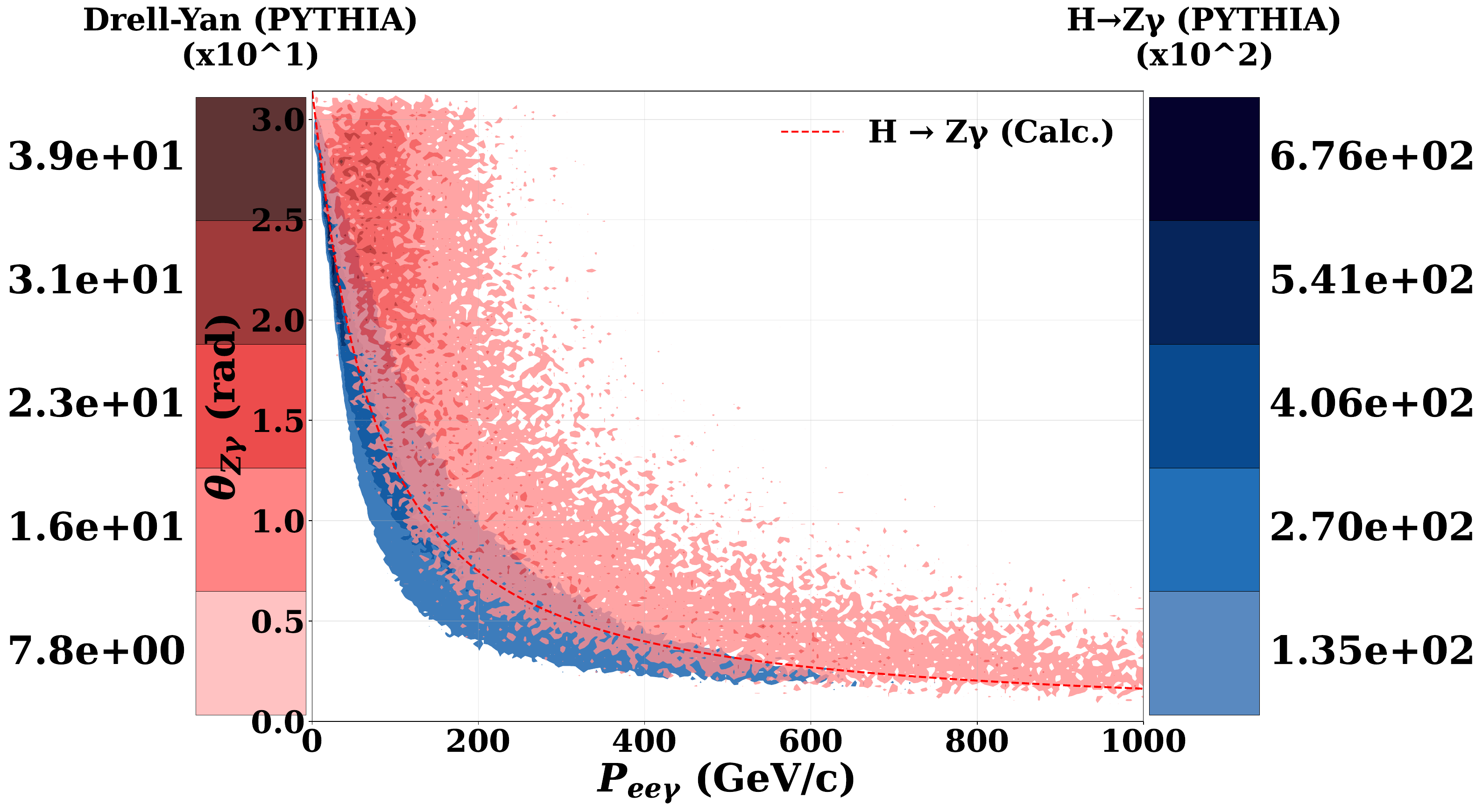}

    \end{minipage}
     \hspace{0.00cm}
    \begin{minipage}{0.48\textwidth}
        \centering
        \includegraphics[width=\textwidth]{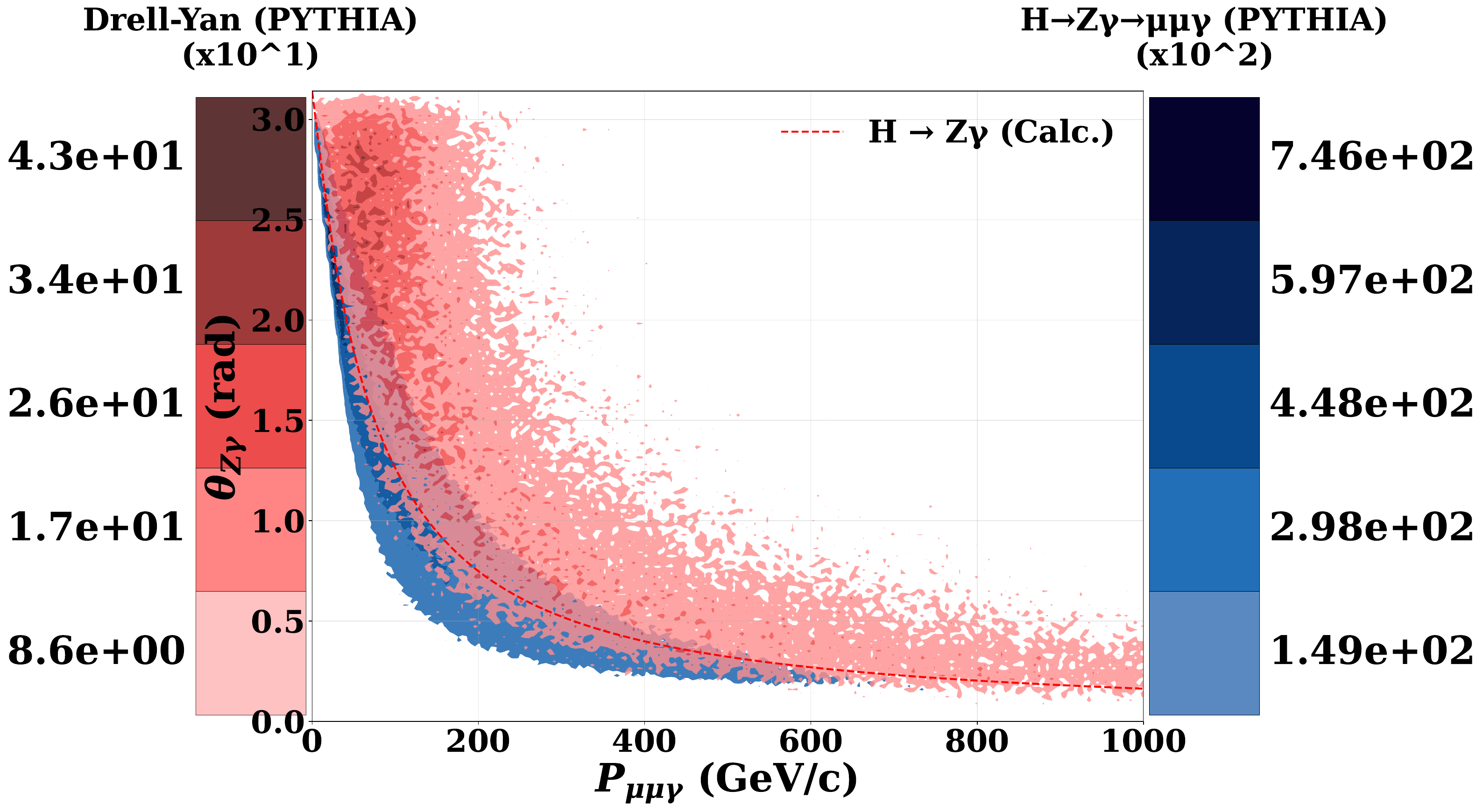}

    \end{minipage}

    \caption{(color online)Kinematic distributions in the $(\theta_{Z\gamma}, P_{\mathrm{Higgs}})$ plane for the electron (left) and muon (right) channels, obtained from Monte Carlo samples. Signal events (blue) exhibit a characteristic correlated structure, while background events (red) populate a broader region without any kinematic constraints.
        }
    \label{fig:2d_plot}
\end{figure}

The analytic model has been validated using Monte Carlo-simulated data, as presented in section~\ref{sec:simulated_samples}.  
Identical event selection and rejection criteria described in Section~\ref{sec:event_selection} are applied to both signal and background samples to construct corresponding two-dimensional distributions in the same phase space. The Drell-Yan process $Z/\gamma^{*} \to \ell\ell$ follows the same angular-momentum correlation as the leptonic decay of the $Z$ boson originating from Higgs decays, since the underlying $Z\to\ell\ell$ topology is identical. However, in radiative $Z/\gamma^{*} + \gamma$ background events, the additional photon is produced independently and does not originate from the same Higgs resonance and is, therefore, not correlated with the momentum of a common parent particle. This difference is assessed by introducing a two-dimensional phase space defined by the opening angle $\theta_{Z\gamma}$ between the di-lepton system and the photon, together with the reconstructed momentum $P_{\mathrm{Higgs}}$ of the $\ell^{+}\ell^{-}\gamma$ system. The signal events from $H\to Z\gamma$ in the PYTHIA8 Monte Carlo samples exhibit a trend in the 2D $(\theta_{Z\gamma}, P_{\mathrm{Higgs}})$ distribution that is consistent with the analytic decay calculation refer as (calc.) in the figure~\ref{fig:2d_plot}. This agreement outlines a well-defined and constrained structure for the signal process in this plane. In contrast, the background events populate the same phase space more broadly, reflecting the absence of Higgs-induced kinematic constraints. To further validate the modeling of angular correlations, we perform an additional study using CMS Ultra-Legacy (UL16) Monte Carlo samples, where the signal is generated with POWHEG+PYTHIA8 and the background with aMC@NLO+PYTHIA8. These generators include matrix-element-level information and are widely used in experimental analyses. The two-dimensional distribution of $\theta_{Z\gamma}$ as a function of $P_{\mathrm{Higgs}}$ is re-evaluated in from these samples. We observe the same characteristic kinematic behaviour as in the baseline study: the signal forms a narrow, well-defined band consistent with a two-body decay topology, whereas the Drell–Yan background exhibits a broader distribution with weaker correlations. This confirms that the relevant angular correlations are preserved in a more realistic and experimentally validated Pythia. These differences further illustrate the distinct kinematic behaviour of signal and background events. The broader background population with weaker correlations compared to the correlated signal structure confirms the effectiveness of the correlated observables in developing discriminating variables between the two processes. Based on these observations, this analysis develops a comprehensive strategy to leverage the distinct kinematic behaviour of the signal and background in the $(\theta_{Z\gamma}, P_{\mathrm{Higgs}})$ plane. Two complementary approaches are pursued: (1) Machine learning techniques that extract discrimination power from correlated observables derived from this phase space and (2) Physics-motivated kinematic selections that directly leverage the observed correlations for background rejection.

The multivariate analysis presented in Section~\ref{sec:kinematic_mva}, employs an XGBoost-based Boosted Decision Tree (BDT) classifier trained on two feature sets: a baseline set of commonly used kinematic variables and an enhanced set that incorporates newly developed correlation based observables derived from the $(\theta_{Z\gamma}, P_{\mathrm{Higgs}})$ plane into these baseline variables. The feature importance rankings, ROC/AUC performance metrics and over-training diagnostics demonstrate that the newly developed observables provide significantly enhanced signal-background separation compared to the baseline variables alone. Section~\ref{sec:Signal-to-Background} quantifies the impact of physics-motivated kinematic selections on signal purity through bin-by-bin analysis of the $(S+B)/B$ ratio in the reconstructed $m_{\ell\ell\gamma}$ spectrum. The momentum-dependent background rejection strategy, constructed by fitting background-dominated regions in the $(\theta_{Z\gamma}, P_{\mathrm{Higgs}})$ plane and defining asymmetric rejection bands, achieves measurable improvements in signal purity while maintaining high signal efficiency. Together, these complementary approaches provide a coherent framework demonstrating the discriminating power and practical applicability of correlation-based observables for Higgs boson analyses in high-luminosity LHC conditions. Section~\ref{sec:summary} summarises the key findings and their implications for future searches.

%======================
\section{Signal and Background Simulation Samples}
\label{sec:simulated_samples}

Monte Carlo (MC) samples for both signal and background processes are generated using \textsc{Pythia8}, which provides a consistent leading-order description of the hard-scattering matrix elements, supplemented with parton showering, initial and final-state radiation, and hadronisation~\cite{Sjostrand2015}. Final-state QED radiation (FSR) is included at the generation level in order to model photon emission from charged leptons. Its impact on the signal selection is subsequently mitigated through dedicated kinematic requirements designed to isolate the genuine $H\rightarrow Z\gamma$ decay topology.
The dominant background contribution originates from the Drell-Yan process $Z/\gamma^{*} \rightarrow \ell^{+}\ell^{-}$. To model this background accurately, $qq \to Z/\gamma^{*}$ events are generated within the di-lepton invariant mass window $60 < m_{\ell\ell} < 120~\mathrm{GeV}$, corresponding to the $Z$-boson resonance region. Separate samples are produced for the di-electron and di-muon final states to account for channel-specific reconstruction effects. Signal samples for the process $H \rightarrow Z\gamma \rightarrow \ell^{+}\ell^{-}\gamma$ are generated inclusively, considering electron and muon decay channels independently. Higgs boson production is simulated via the gluon--gluon fusion (ggF) mechanism, which dominates the inclusive Higgs production rate at a centre-of-mass energy of $\sqrt{s}=13~\mathrm{TeV}$. The corresponding production cross section is approximately $\sigma_{\mathrm{ggF}} \simeq 48.6~\mathrm{pb}$ for a Higgs boson mass of $m_{H}=125~\mathrm{GeV}$, as predicted by perturbative QCD calculations at next-to-next-to-leading order and higher~\cite{LHCHiggsXS}. This production mode, therefore, provides the primary contribution to the expected signal yield.

\section{Event Selection}
\label{sec:event_selection}

A minimal yet robust event selection is applied to define a clean and well-controlled phase space for signal and background samples. Events are required to contain exactly two same-flavour, opposite-sign leptons and one photon, consistent with the targeted final state. In the di-electron channel, the leading and subleading electrons from the $Z$-boson decay must satisfy $p_{T}(e_{\mathrm{lead}}) > 25~\mathrm{GeV}$ and $p_{T}(e_{\mathrm{sublead}}) > 15~\mathrm{GeV}$, respectively. In the di-muon channel, the corresponding thresholds are $p_{T}(\mu_{\mathrm{lead}}) > 20~\mathrm{GeV}$ and $p_{T}(\mu_{\mathrm{sublead}}) > 10~\mathrm{GeV}$. The reconstructed photon is required to have $p_{T}(\gamma) > 15~\mathrm{GeV}$ in both channels. To suppress final-state radiation and collinear configurations, additional kinematic constraints are imposed. The invariant mass of the di-lepton system is required to satisfy $Z_{M} > 50~\mathrm{GeV}$. The angular separation between each lepton and the photon must satisfy $\Delta R(\ell_{1},\gamma) > 0.4$ as well as $\Delta R(\ell_{2},\gamma) > 0.4$ for both decay channels. These requirements efficiently reject topologies dominated by photon emission from final-state leptons. To enhance consistency with the $H \rightarrow Z\gamma$ decay hypothesis, the sum of the reconstructed Higgs candidate mass and the di-lepton invariant mass is required to satisfy $(M_{H} + Z_{M}) > 185~\mathrm{GeV}$~\cite{Hzgamma_cms}. This selection reduces residual background while preserving the signal's characteristic kinematic features. The resulting event sample contains two well-identified isolated leptons and a high-energy photon, providing a suitable basis for subsequent signal extraction.

\section{Experimental Analyses vs Present Approach}

To place our study in context, we briefly review the strategies used in ATLAS and CMS analyses of the $H \to Z\gamma$ channel and clarify how our approach differs from these experimental methods. In these experimental studies, multivariate techniques such as boosted decision trees are primarily employed for event categorisation, where events are divided into multiple exclusive categories based on production modes and kinematic properties in order to enhance the overall sensitivity. The final signal extraction is then performed through a simultaneous fit to the invariant mass distribution of the $\ell^+\ell^-\gamma$ system across all categories. For instance, the CMS analysis using the full Run-2 dataset of 138 fb$^{-1}$ follows this strategy, where multivariate discriminants are used to define categories, and the signal is subsequently extracted from a likelihood fit to the invariant mass spectrum~\cite{Hzgamma_cms, Hzgamma_atlas, Hzgamma_atlas_run3}. In contrast, the our study adopts a different strategy. Rather than using machine learning for event categorisation, we employ it directly as a classifier to distinguish signal from background (Drell-Yan) events in a common analysis phase space. This allows us to focus on the intrinsic discriminating power of kinematic observables without introducing additional complexity from category-based optimisations. Furthermore, our study considers a simplified scenario with a single dominant (Drell-Yan) background, and therefore does not perform event categorisation or a full statistical signal extraction based on invariant mass fits. Consequently, a direct comparison with experimental results in terms of signal significance or invariant mass fit performance is not appropriate, as the two approaches are conceptually different and operate at different levels of analysis. Instead, we present complementary studies based on machine learning classification and signal-to-background ($S+B/B$) ratios to quantify the improvement in discrimination arising from the inclusion of correlation-sensitive observables. This approach enables a controlled assessment of the additional physics information captured by the proposed variables, independent of the categorisation strategies used in experimental analyses.

\section{Kinematic MVA Training}
\label{sec:kinematic_mva}

\subsection{Event Pre-processing and Input Variables}

\label{subsec:event_preproc_traning_var}
Machine learning techniques are widely used in high-energy physics for event reconstruction, classification, and analysis optimization~\cite{ml_hep1, ml_hep2, ml_hep3, ml_hep4}. In this work, discrimination of signal from background events using an XGBoost-based Boosted Decision Tree, with performance evaluated through ROC/AUC metrics and over-training diagnostics~\cite{Hocker2007TMVA_overtraining_bdt_auc_roc}. Events passing the selection described in Section~\ref{sec:event_selection} are used for the multivariate analysis. By applying identical selection requirements to both signal and background MC samples, a common analysis phase space is established, ensuring that any differences learned by the classifier are due to genuine physical differences in event kinematics, rather than artefacts of the selection or detector acceptance. The input variables for the multivariate training are summarised in Table~\ref{tab:inputvars}, which compares the two training configurations used in this study. Training Set--1 (baseline) includes nine standard kinematic and angular observables commonly used in the previous studies: the separation between leptons $\Delta R_{\ell,\ell}$, pseudorapidities $\eta_{\ell_1}$, $\eta_{\ell_2}$, $\eta_{\gamma}$, the polar angle ($\cos\theta$), azimuthal angles, $\phi$ in the $Z$-boson rest frame, the ratio $p_{T}^{\ell\ell\gamma}/m_{\ell\ell\gamma}$, the di-lepton invariant mass $m_{\ell\ell}$ and the photon transverse momentum $p_{T}(\gamma)$~\cite{add_angular1, add_angular2, add_angular3}. Although these variables encode the global kinematic properties of the event, their sensitivity to the correlation structure arising from the two-body decay kinematics of the Higgs boson remains limited.

To address this, Training Set--2 extends the baseline by including three additional correlated observables: the opening angle $\theta_{Z\gamma}$ between the $Z$ boson and the photon, the reconstructed Higgs momentum $P_{\mathrm{Higgs}}$ and the composite variable $\log(\theta_{Z\gamma} \times P_{\mathrm{Higgs}})$. These variables are specifically designed to make use of the characteristic kinematic correlations present in signal events, as discussed in Section~\ref{sec:intro} and provide enhanced discrimination power against the Drell-Yan background. The inclusion of these observables allows the multivariate classifier to access information about the interplay between angular and momentum variables that is not available in the baseline set alone. The combination of standard and correlation-based observables ensures that the analysis is comprehensive and sensitive to the unique features of $H\rightarrow Z\gamma$ signal events.

\begin{table}[H]
\centering
\footnotesize
\renewcommand{\arraystretch}{1.25}
\begin{tabular}{|l|c|c|l|}
\hline
\textbf{Variable Name} & \textbf{Set--1} & \textbf{Set--2} & \textbf{Type} \\
\hline
$\Delta R_{\ell,\ell}$ & $\checkmark$ & $\checkmark$ & Angular \\
$\eta_{\ell_{1}}$     & $\checkmark$ & $\checkmark$ & Angular \\
$\eta_{\ell_{2}}$     & $\checkmark$ & $\checkmark$ & Angular \\
$\eta_{\gamma}$       & $\checkmark$ & $\checkmark$ & Angular \\
$\cos\theta$          & $\checkmark$ & $\checkmark$ & Angular \\
$\phi$                & $\checkmark$ & $\checkmark$ & Angular \\
\hline
$p_{T}^{\ell\ell\gamma}/m_{\ell\ell\gamma}$ 
                      & $\checkmark$ & $\checkmark$ & Boosted Higgs \\
\hline
$m_{\ell\ell}$        & $\checkmark$ & $\checkmark$ & Mass \\
\hline
$p_{T}(\gamma)$       & $\checkmark$ & $\checkmark$ & Photon $p_{T}$ \\
\hline
$\theta_{Z\gamma}$    & --            & $\checkmark$ & Angular \\
$P_{\mathrm{Higgs}}$  & --            & $\checkmark$ & Momentum \\
$log(\theta_{Z\gamma} \times P_{\mathrm{Higgs}})$ 
                      & --            & $\checkmark$ & Correlated variables \\
\hline
\end{tabular}
\caption{List of input observables used in the multivariate analysis. Training Set--1 corresponds to the baseline configuration, while Training Set--2 extends the baseline with additional correlated observables.}
\label{tab:inputvars}
\end{table}

\subsection{Pre-training Variable Separation and KS Test Evaluation}
\label{subsec:pre_trainng_var_sep_ks}

\begin{figure}[hbt!]
\centering
    \includegraphics[width=1.0\textwidth]{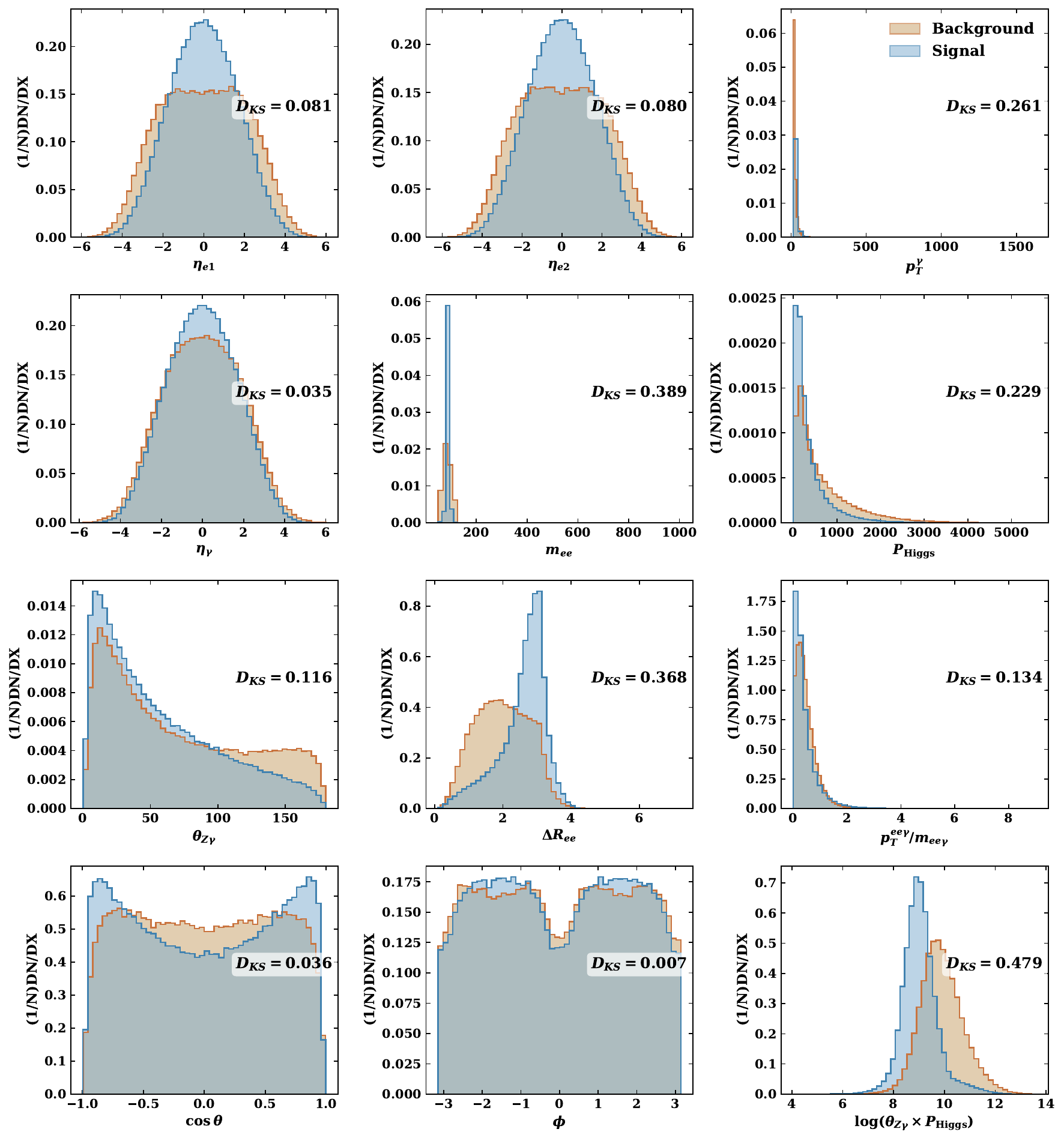}
 \caption{(color online)Overlaid signal and background distributions in the case of electrons for all-purpose variables. The Kolmogorov--Smirnov statistic $D_{\mathrm{KS}}$ is calculated to quantify the shape differences between the signal and background distributions.}

\label{fig:inputvars_sep_electron}
\end{figure}

\begin{figure}[hbt!]
\centering
    \includegraphics[width=1.0\textwidth]{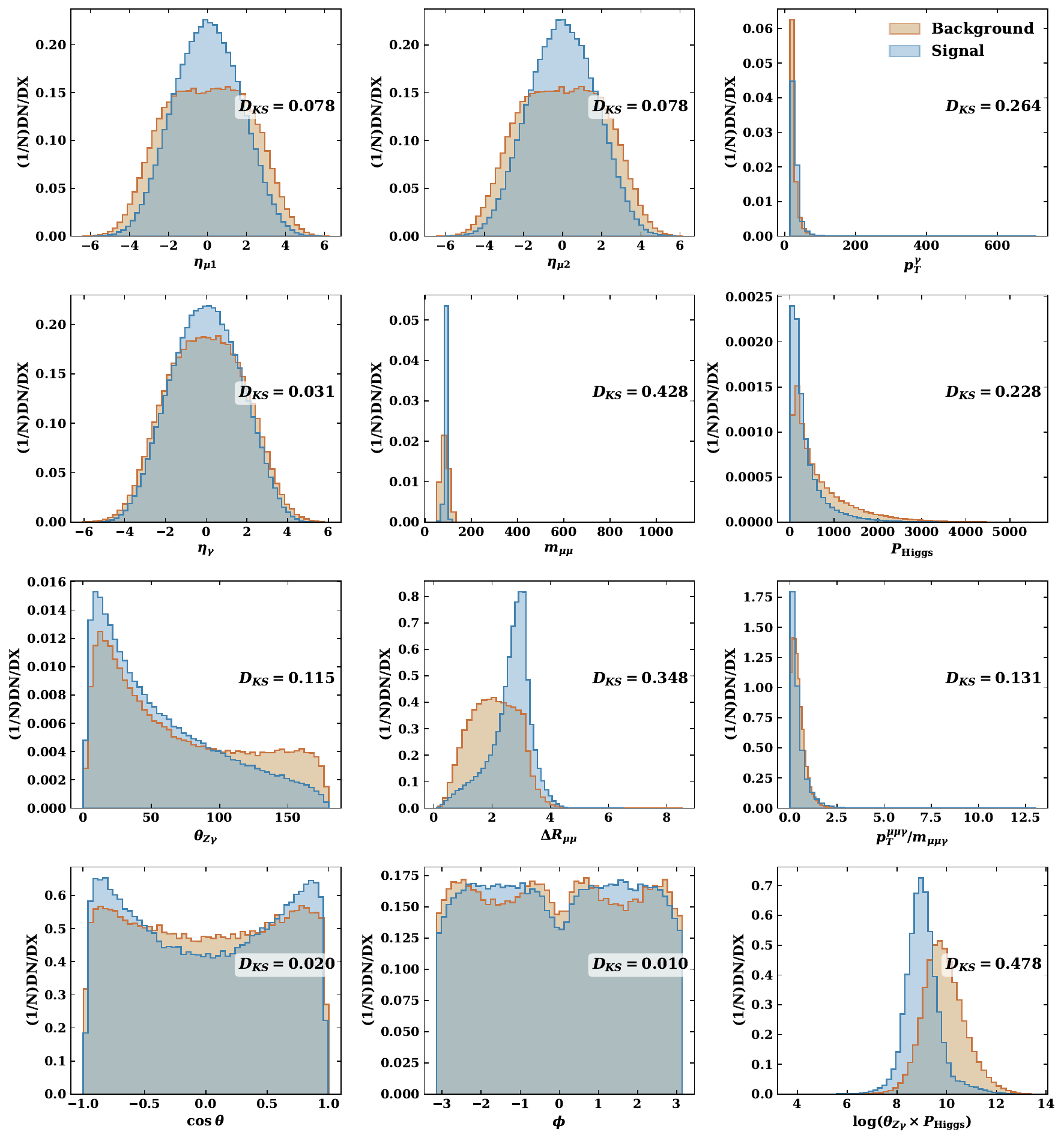}
 \caption{(color online)Overlaid signal and background distributions in case of muons for the all-purpose variables. The Kolmogorov--Smirnov statistic $D_{\mathrm{KS}}$ is calculated to quantify the shape difference between the signal and background distributions.}
\label{fig:inputvars_sep_muon}
\end{figure}

The separation between signal and background for the input observables is evaluated before performing the full multivariate training, serving as a valuable cross-check of the physical intuition underlying the choice of input observables. For each variable, the one-dimensional distributions of the signal and background are examined to assess differences in shape and population. In addition to a visual inspection of these distributions, the statistical separation is quantified by $D_\mathrm{KS}$ using the two-sample~\cite {Kolmogorov1933_ks, Smirnov1948_ks, Massey1951_ks}. For each observable, the KS test statistic is computed, which measures the maximum vertical distance between the cumulative distribution functions (CDFs) of the two samples and provides a direct shape-based measure of how different the signal and background distributions are. A higher value of $D_\mathrm{KS}$ indicates a stronger separation power of the corresponding input variable, as their distributions are well separated. A low value suggests significant overlap and limited discriminating power. The variable $log(\theta_{Z\gamma} × P_{Higgs})$ with the highest $D_{\mathrm{KS}}=$0.497(electrons) and 0.478(muons) is prioritized for inclusion in the enhanced feature set, as it provides the most significant potential to improve signal-background separation in subsequent BDT training presented in Figure~\ref{fig:inputvars_sep_electron} and~\ref{fig:inputvars_sep_muon}. 

Among all input observables, the correlated variable $\log(\theta_{Z\gamma} \times P_{\mathrm{Higgs}})$ exhibits one of the strongest separations. Signal and $Z/\gamma^{*}$ background events populate clearly distinct regions of this observable, as shown in Figures~\ref{fig:inputvars_sep_electron} and~\ref{fig:inputvars_sep_muon}, confirming its additional discriminating power between signal and background. The achieved separation further motivates its inclusion as an input to the multivariate training.

\subsection{Feature Importance Ranking}
\label{subsec:features_importance}
The relative contributions of the input variables to the multivariate classifier are quantified using the XGBoost feature-importance metric. The variable $log(\theta_{Z\gamma} \times P_H)$ exhibits the highest importance, with typical values of 0.233 and 0.221 for the electrons and muons, respectively, significantly larger than other angular observables, again indicating that this combined variable carries the largest fraction of the information used by the classifier to separate signal from the background. Their product, combined with the logarithmic transformation, enhances sensitivity by compressing the wide kinematic range and emphasising regions where the signal-background separation is optimal. Table~\ref{tab:featureimportance_tabel} includes the full feature-importance ranking to illustrate the relative impact of all input variables and to highlight the dominant role played by the newly introduced observables $\log(\theta_{Z\gamma} \times P_H)$ for both electrons and muons.

\begin{table}[H]
\centering
\footnotesize
\renewcommand{\arraystretch}{1.25}

\begin{minipage}{0.48\textwidth}
\centering
\vspace{0.2cm}
\begin{tabular}{|l|c|}
\hline
\textbf{Variable} & \textbf{Importance} \\
\hline
$m_{ee}$                                   & $3.70 \times 10^{-1}$ \\
$\log(\theta_{Z\gamma} \times P_{\mathrm{Higgs}})$   & $2.33 \times 10^{-1}$ \\
$p_{T}^{\gamma}$                            & $9.86 \times 10^{-2}$ \\
$\Delta R_{ee}$                         & $8.67 \times 10^{-2}$ \\
$P_{\mathrm{Higgs}}$                             & $4.66 \times 10^{-2}$ \\
$\theta_{Z\gamma}$                        & $4.22 \times 10^{-2}$ \\
$p_{T}^{ee\gamma}/m_{ee\gamma}$                      & $3.97 \times 10^{-2}$ \\
$\eta_{\gamma}$                           & $2.99 \times 10^{-2}$ \\
$\eta_{e_2}$                             & $1.55 \times 10^{-2}$ \\
$\eta_{e_1}$                             & $1.50 \times 10^{-2}$ \\
$\cos\theta_{1}^{e}$                    & $1.22 \times 10^{-2}$ \\
$\phi_{1}^{e}$                         & $1.02 \times 10^{-2}$ \\
\hline
\end{tabular}
\end{minipage}
\hfill
\begin{minipage}{0.48\textwidth}
\centering
\vspace{0.2cm}
\begin{tabular}{|l|c|}
\hline
\textbf{Variable} & \textbf{Importance} \\
\hline
$m_{\mu\mu}$                                   & $4.02 \times 10^{-1}$ \\
$\log(\theta_{Z\gamma} \times P_{\mathrm{Higgs}})$   & $2.21 \times 10^{-1}$ \\
$p_{T}^{\gamma}$                            & $9.85 \times 10^{-2}$ \\
$\Delta R_{\mu\mu}$                         & $7.55 \times 10^{-2}$ \\
$P_{\mathrm{Higgs}}$                             & $4.64 \times 10^{-2}$ \\
$\theta_{Z\gamma}$                        & $4.15 \times 10^{-2}$ \\
$p_{T}^{\mu\mu\gamma}/m_{\mu\mu\gamma}$                      & $3.87 \times 10^{-2}$ \\
$\eta_{\gamma}$                           & $2.83 \times 10^{-2}$ \\
$\eta_{\mu_2}$                             & $1.38 \times 10^{-2}$ \\
$\eta_{\mu_1}$                             & $1.32 \times 10^{-2}$ \\
$\cos\theta_{1}^{\mu}$                    & $1.11 \times 10^{-2}$ \\
$\phi_{1}^{\mu}$                         & $9.39 \times 10^{-3}$ \\
\hline
\end{tabular}
\end{minipage}

\caption{Feature-importance values from the XGBoost classifier for the electron (left) and muon (right).}
\label{tab:featureimportance_tabel}
\end{table}

\subsection{Training Setup and Classifier Performance}
\label{subsec:training_setup}
The classifier is based on gradient boosting techniques, and implemented using the XGBoost algorithm~\cite{xgboost1,xgboost2}, which is trained to separate the $H\!\to Z\gamma$ signal from the $Z/\gamma^*$ background.  
Signal and background samples are divided into statistically independent training, validation, and testing sub-samples using a fixed splitting fraction to ensure unbiased performance evaluation and to monitor over-training. The validation set is used to tune hyperparameters and prevent overfitting, while the test set provides an unbiased final estimate of model performance~\cite{scikit-learn}. The model has been trained using the binary logistic loss function with hyperparameters such as the tree depth~\cite{cox1958regression}, learning rate, and number of estimators, which are optimised to achieve stable convergence. All input observables are normalised before training. This ensures that each feature contributes equally to the learning process, preventing features with larger scales from dominating the optimisation. Two BDT classifiers are considered: a classifier corresponding to set--1 input variables and a classifier corresponding to set--2 variables via incorporating additional correlated observables as described in Section~\ref{subsec:event_preproc_traning_var}, hereafter referred to as the \emph{baseline} and \emph{extended} models. The performance of the two classifiers is evaluated and presented in the following subsections.

\subsubsection{ROC/AUC, BDT Output Score and Efficiencies}
\begin{figure}[H]
\centering
% ----------- First row: Signal plots -----------
\begin{minipage}{0.48\textwidth}
    \centering
    \includegraphics[width=\textwidth]{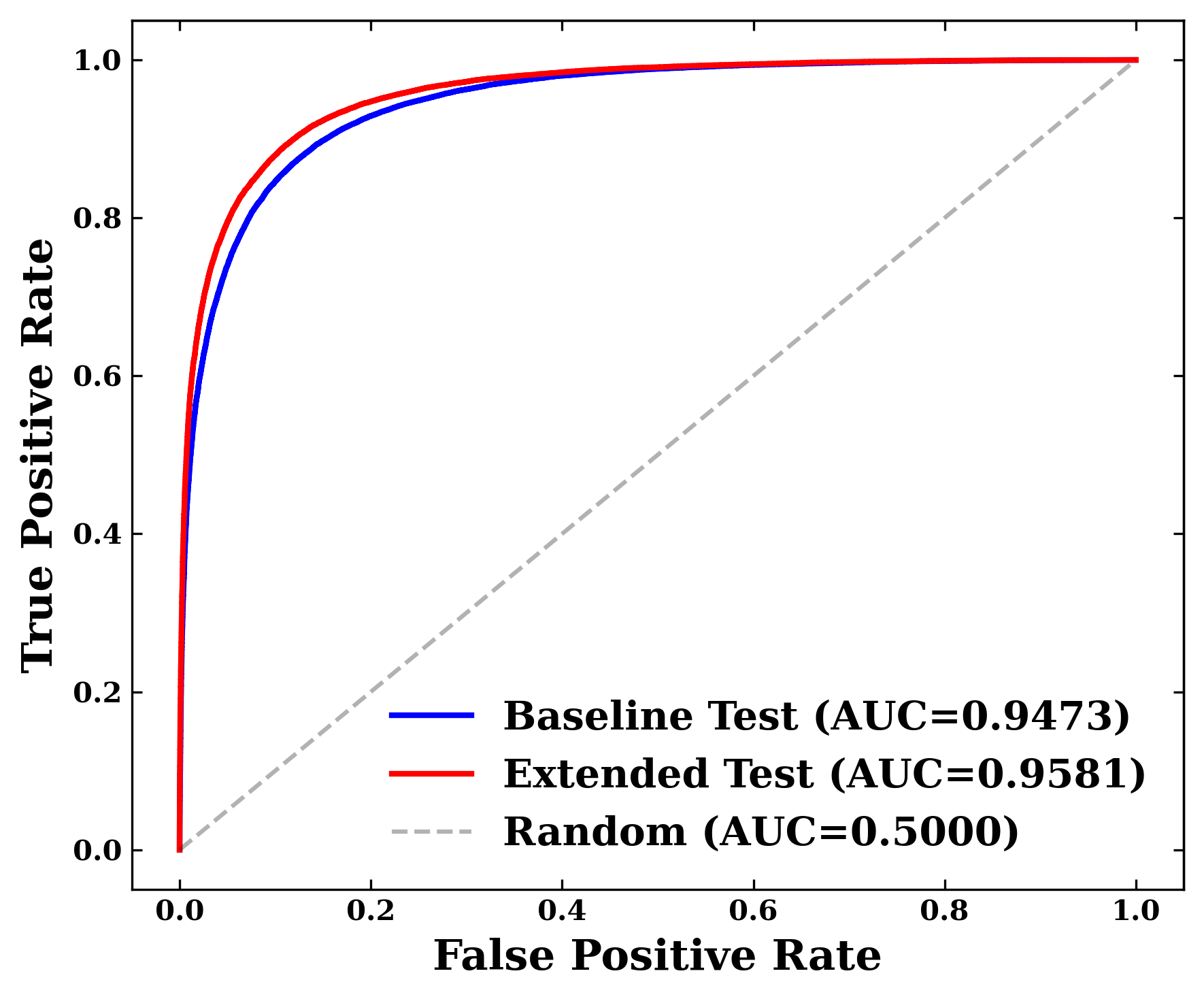}
\end{minipage}
\hfill
\begin{minipage}{0.48\textwidth}
    \centering
    \includegraphics[width=\textwidth]{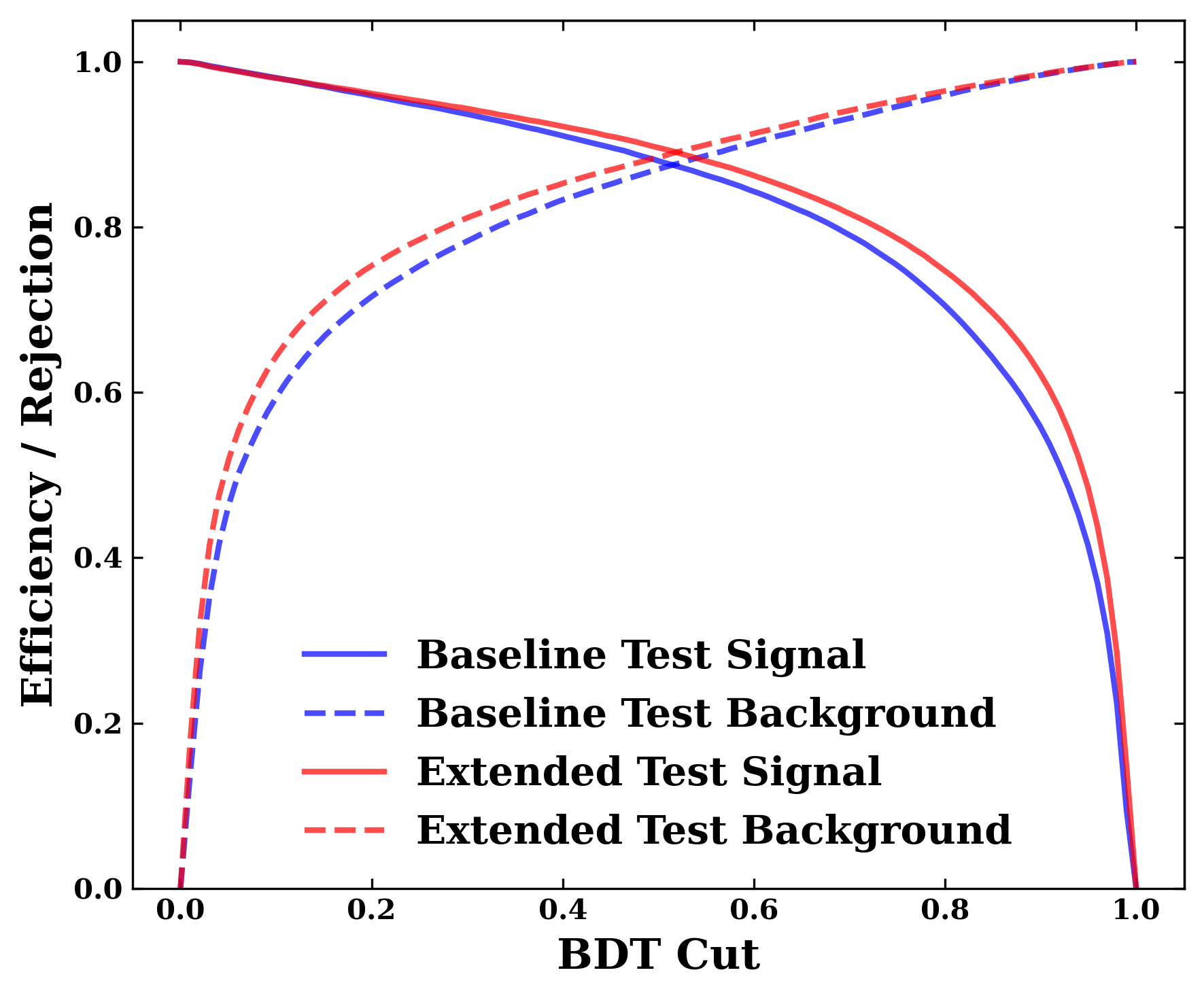}
\end{minipage}

\vspace{4pt}

% ----------- Second row: Background plots -----------
\begin{minipage}{0.48\textwidth}
    \centering
    \includegraphics[width=\textwidth]{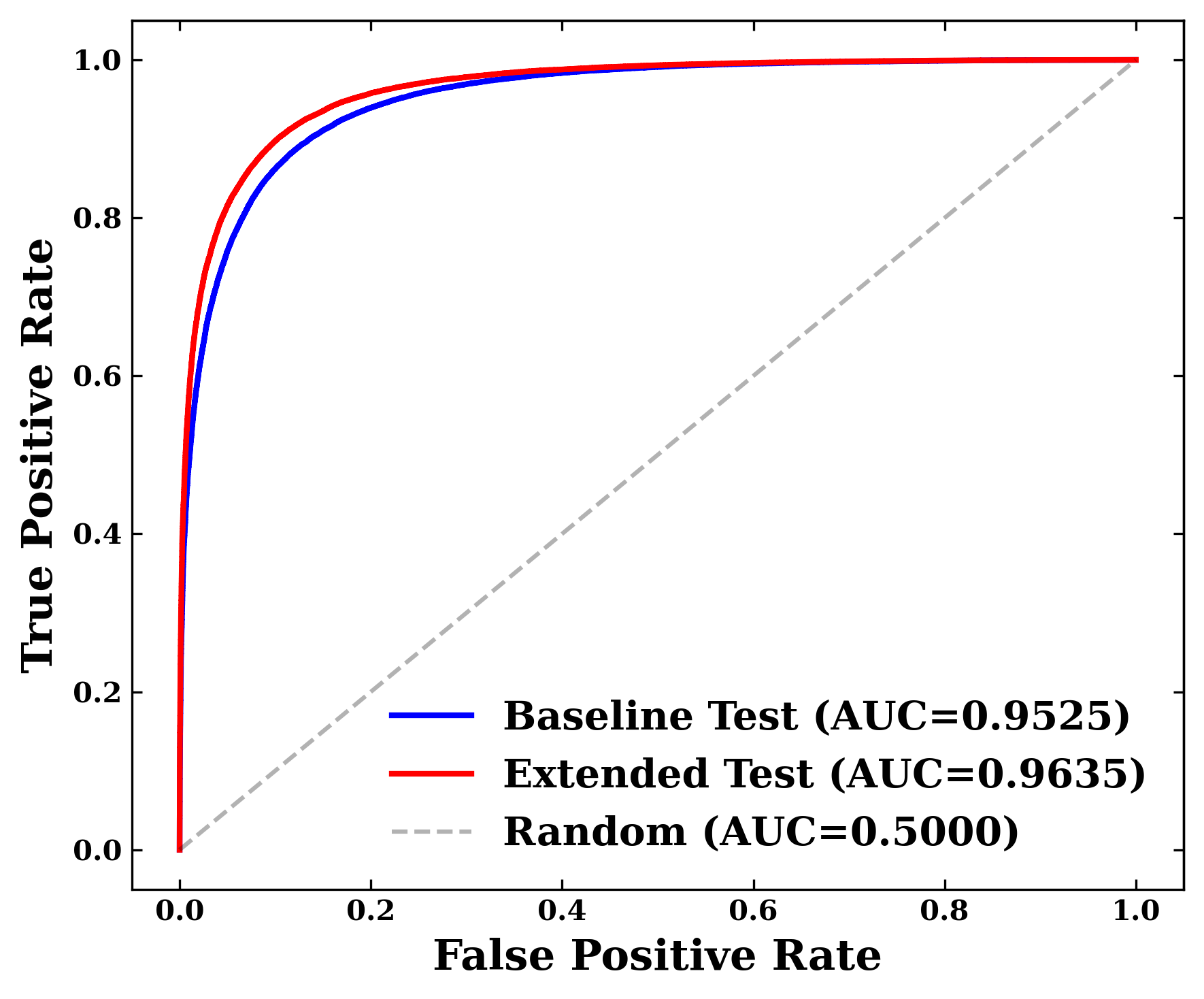}
\end{minipage}
\hfill
\begin{minipage}{0.48\textwidth}
    \centering
    \includegraphics[width=\textwidth]{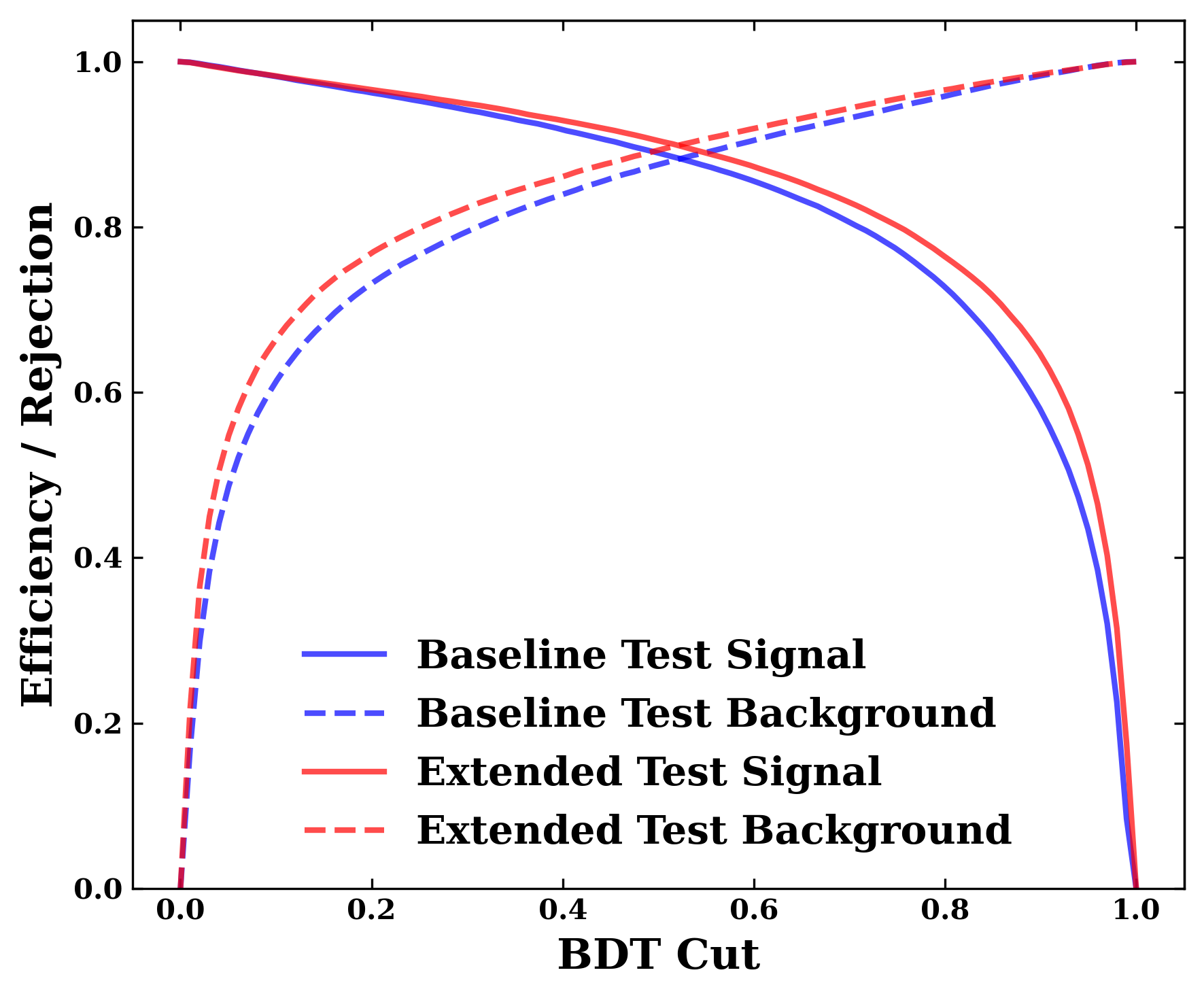}
\end{minipage}

\caption{(color online)ROC curves (\emph{left}) and the corresponding signal and background efficiencies (\emph{right}) for the electron (\emph{top}) and muon (\emph{bottom}) channels to evaluate the performance of the BDT classifiers.}
\label{fig:roc_eff}
\end{figure}

The inclusion of correlation-based observables leads to a clear and consistent improvement in the performance of the \emph{extended} classifier in both lepton channels.
This improvement is quantified by an increase in the area under the ROC curve~\cite{Fawcett2006_Roc2}, with $\Delta\mathrm{AUC}=+0.0108$ in the electron channel (0.9473 $\rightarrow$ 0.9581) and $\Delta\mathrm{AUC}=+0.0110$ in the muon channel (0.9525 $\rightarrow$ 0.9635), as summarized in Table~\ref{tab:auc_overtraining_summary} and illustrated in Figure~\ref{fig:roc_eff}.

\begin{table}[hbt!]
\centering
\footnotesize
\renewcommand{\arraystretch}{1.25}
\begin{tabular}{|c|c|c|c|c|}
\hline
\textbf{Channel} & \textbf{Model} & \textbf{AUC (Train)} & \textbf{AUC (Test)} & 
$\boldsymbol{|\Delta\text{AUC}|_{\text{train--test}}}$ \\
\hline
Electron & Baseline BDT & 0.9562 & 0.9473 & 0.0089 \\
 & Extended BDT & 0.9674 & 0.9581 & 0.0093 \\
\hline
Muon & Baseline BDT & 0.9591 & 0.9525 & 0.0066 \\
 & Extended BDT & 0.9703 & 0.9635 & 0.0068 \\
\hline
\multicolumn{5}{|c|}{$\boldsymbol{\Delta\text{AUC}_{\text{Train}}^{e} = +0.0112 
\qquad \Delta\text{AUC}_{\text{Train}}^{\mu} = +0.0112}$} \\

\multicolumn{5}{|c|}{$\boldsymbol{\Delta\text{AUC}_{\text{Test}}^{e} = +0.0108 
\qquad \Delta\text{AUC}_{\text{Test}}^{\mu} = +0.0110}$} \\
\hline
\end{tabular}
\caption{Training and test AUC values for the \emph{baseline} and \emph{extended} classifiers in the electron and muon channels.}
\label{tab:auc_overtraining_summary}
\end{table}

The efficiency–rejection curves in Figure~\ref{fig:roc_eff} further show that, for a given BDT threshold, the extended model achieves higher signal efficiency at the same background rejection, or equivalently stronger background rejection at fixed signal efficiency.
The improved performance across a wide range of BDT cuts confirms the enhanced discriminating power and robustness provided by the correlated kinematic observables. Although the numerical increase in AUC is modest, it is achieved in a phase-space region dominated by irreducible Drell-Yan backgrounds. It, therefore, represents a non-trivial enhancement in discriminating power. Beyond the global AUC metric, the improvement is also reflected in the shape of the BDT output distributions.

\begin{figure}[H]
\centering

% ----------- First row: Signal plots -----------
\begin{minipage}{0.48\textwidth}
    \centering
    \includegraphics[width=\textwidth]{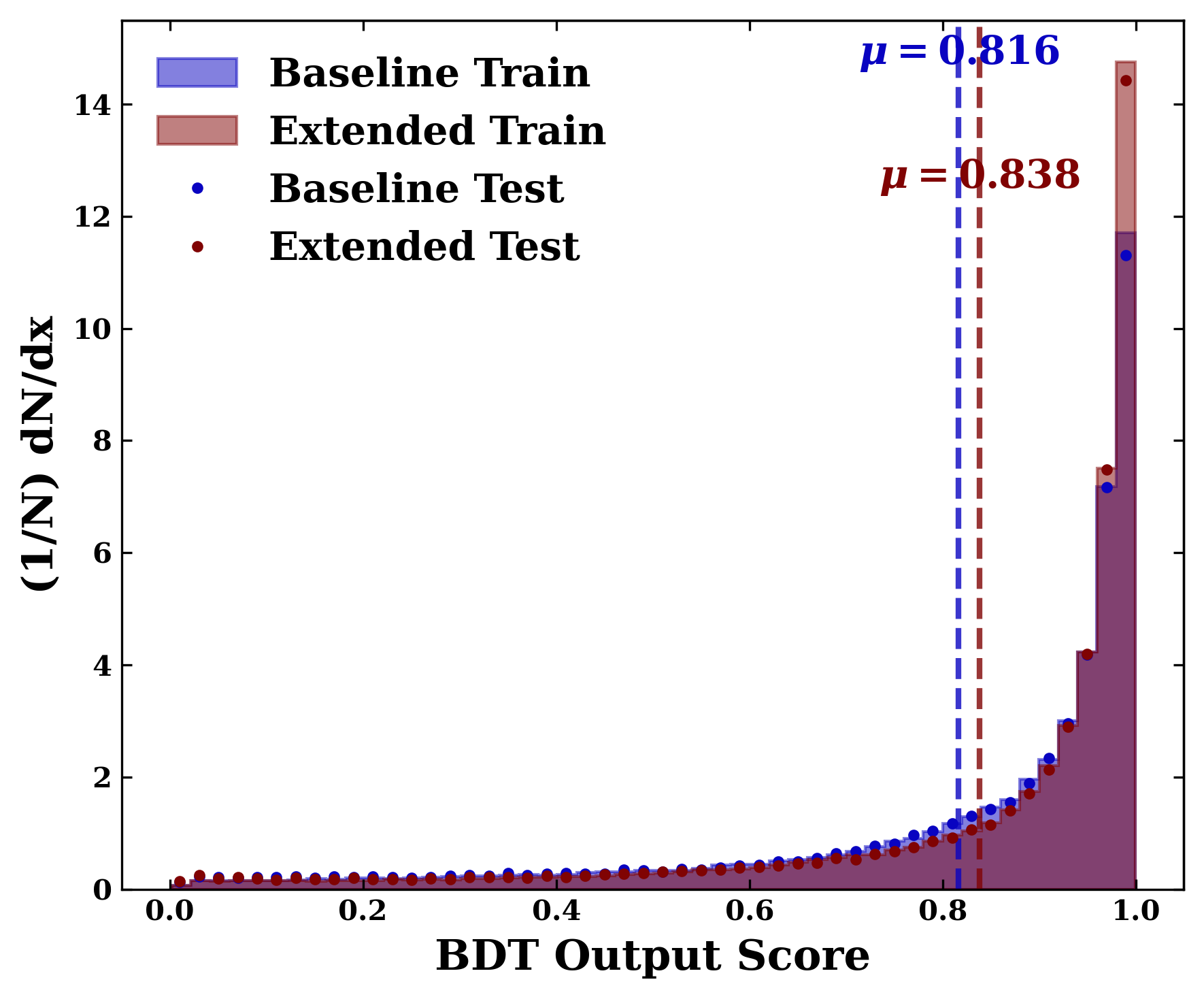}
\end{minipage}
\hfill
\begin{minipage}{0.48\textwidth}
    \centering
    \includegraphics[width=\textwidth]{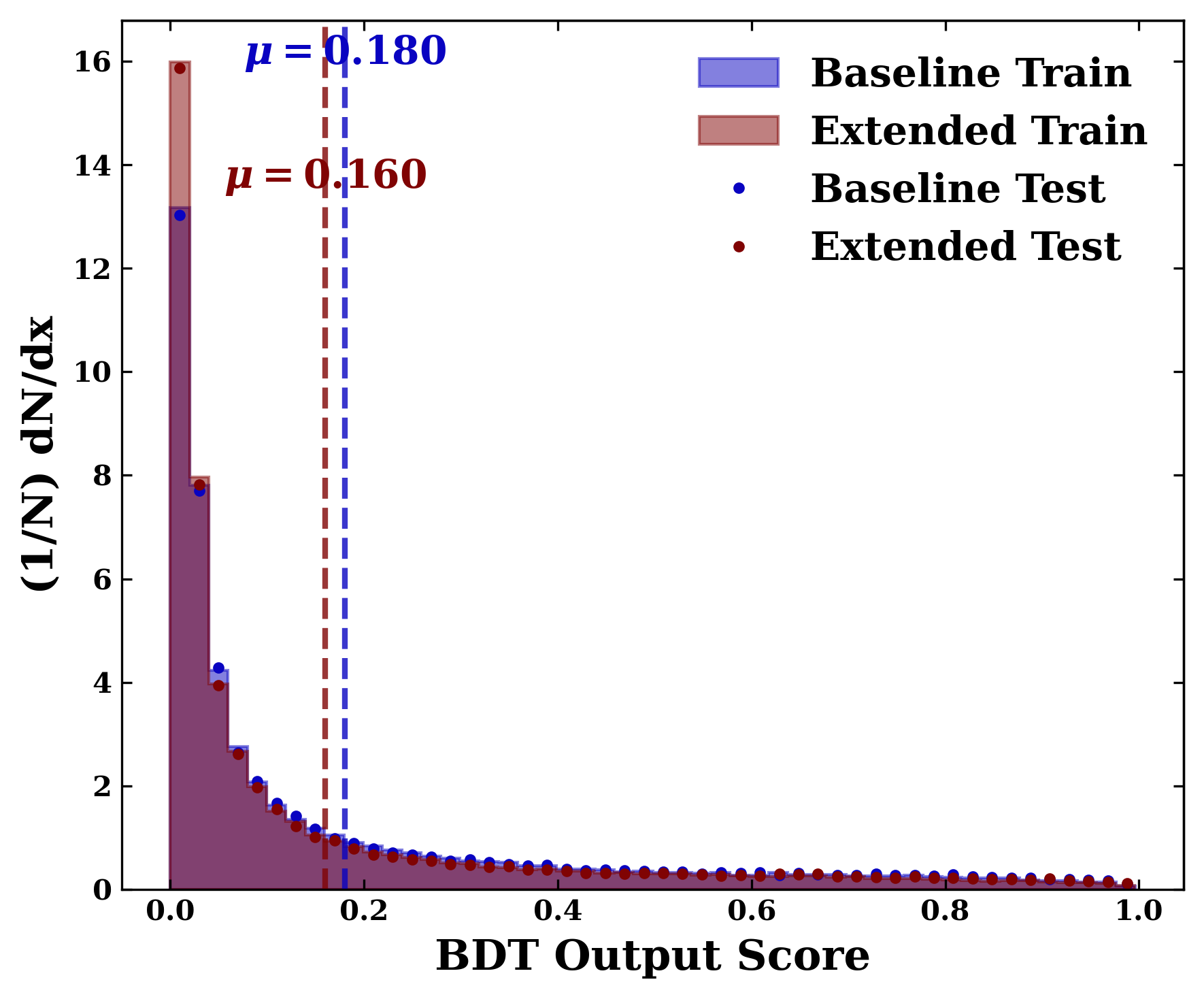}
\end{minipage}

\vspace{4pt}

% ----------- Second row: Background plots -----------
\begin{minipage}{0.48\textwidth}
    \centering
    \includegraphics[width=\textwidth]{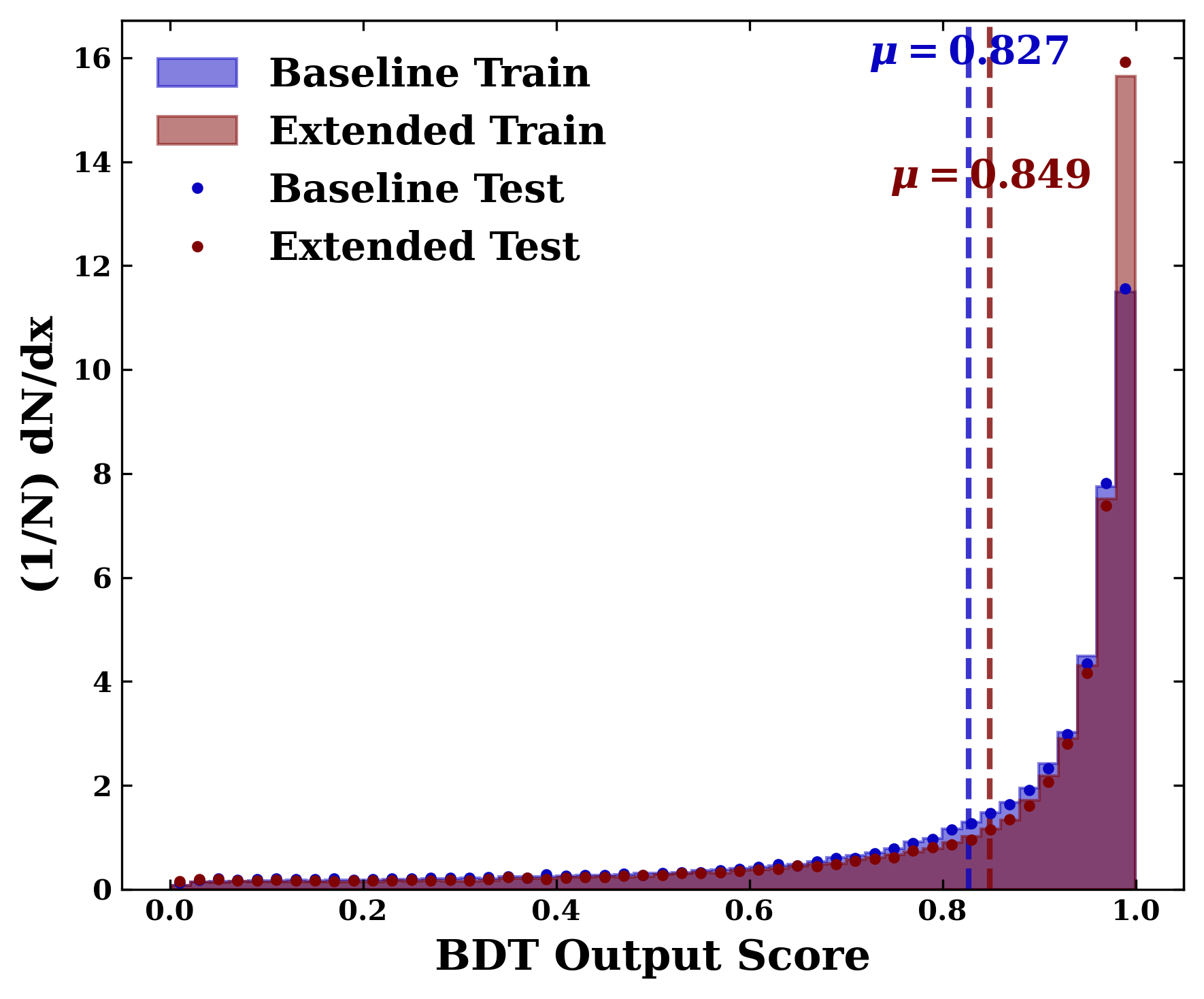}
\end{minipage}
\hfill
\begin{minipage}{0.48\textwidth}
    \centering
    \includegraphics[width=\textwidth]{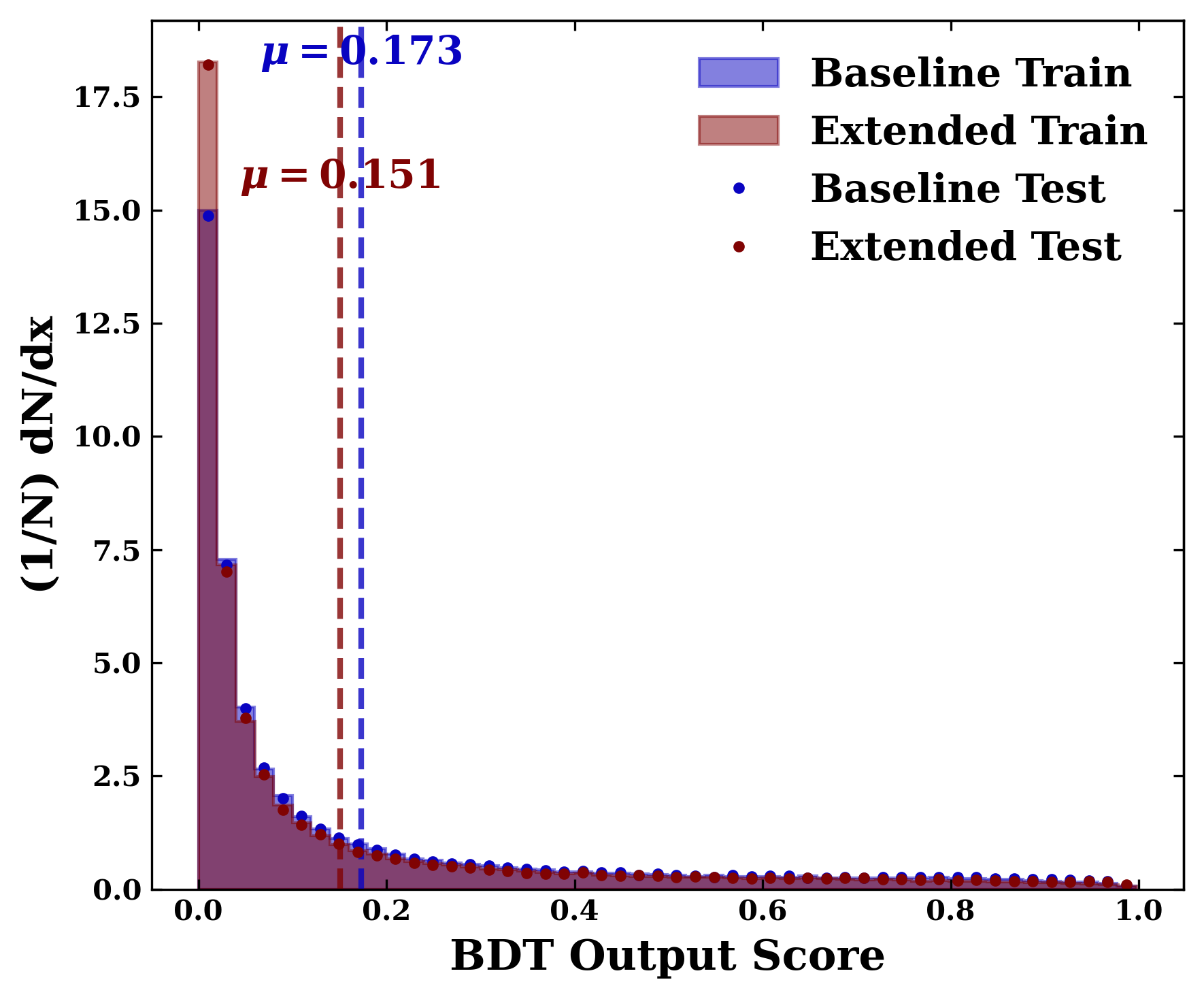}
\end{minipage}

\caption{(color online)BDT classifier performance in the electron channel (\emph{top}) and muon channel (\emph{bottom}), comparing the \emph{baseline} and \emph{extended} models.
The panels display the BDT output score distributions for signal (\emph{left}) and background (\emph{right}) events.}

\label{fig:performance}
\end{figure}

In the electron channel, the mean BDT output for signal events shifts towards higher values $\Delta\mu_{\mathrm{sig}}^{e} > 0$, while the background distribution migrates towards lower values $\Delta\mu_{\mathrm{bkg}}^{e} < 0$. An analogous behaviour is observed in the muon channel, with $\Delta\mu_{\mathrm{sig}}^{\mu} > 0$ and $\Delta\mu_{\mathrm{bkg}}^{\mu} < 0$, (see Figure \ref{fig:performance}). These coherent shifts demonstrate that the additional observables systematically push the signal and background into more distinct regions of the classifier output. The consistent improvement observed across both lepton final states indicates that the performance gain is robust and not driven by channel-specific effects or statistical fluctuations. Instead, it reflects genuine kinematic differences between the resonant $H\to Z\gamma$ signal and the dominant $Z/\gamma^{*}$ background. In particular, the inclusion of the correlated variable $\log(\theta_{Z\gamma} \times P_{\mathrm{H}})$ plays a key role in utilising the momentum-dependent angular structure of the Higgs boson decay, leading to enhanced signal--background separation.

\subsubsection{Performance Gain}

The performance gain of the \emph{extended} classifier relative to the  \emph{baseline} is quantified in terms of changes in the background misidentification rate (FPR) and the signal efficiency (TPR). For a given target signal efficiency, the corresponding background misidentification rates are extracted from the ROC curves for both classifiers and the FPR improvement is defined as $\mathrm{FPR}_{\mathrm{base}} - \mathrm{FPR}_{\mathrm{ext}}$ expressed in percentage. Positive values \emph{(blue)} indicate an enhanced reduction of the background rate at fixed signal efficiency. Conversely, for a fixed background misidentification rate, the signal efficiencies of the two classifiers are compared. 

\begin{figure}[H]
\centering
\includegraphics[width=1.0\textwidth]{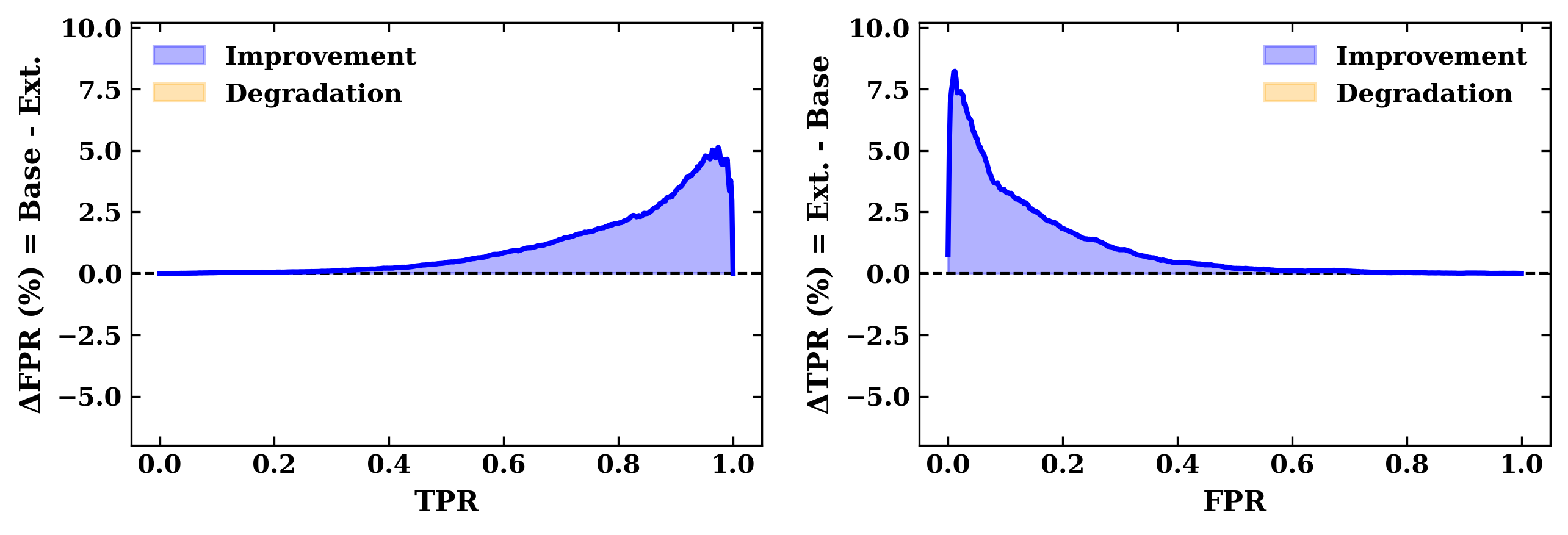}
\includegraphics[width=1.0\textwidth]{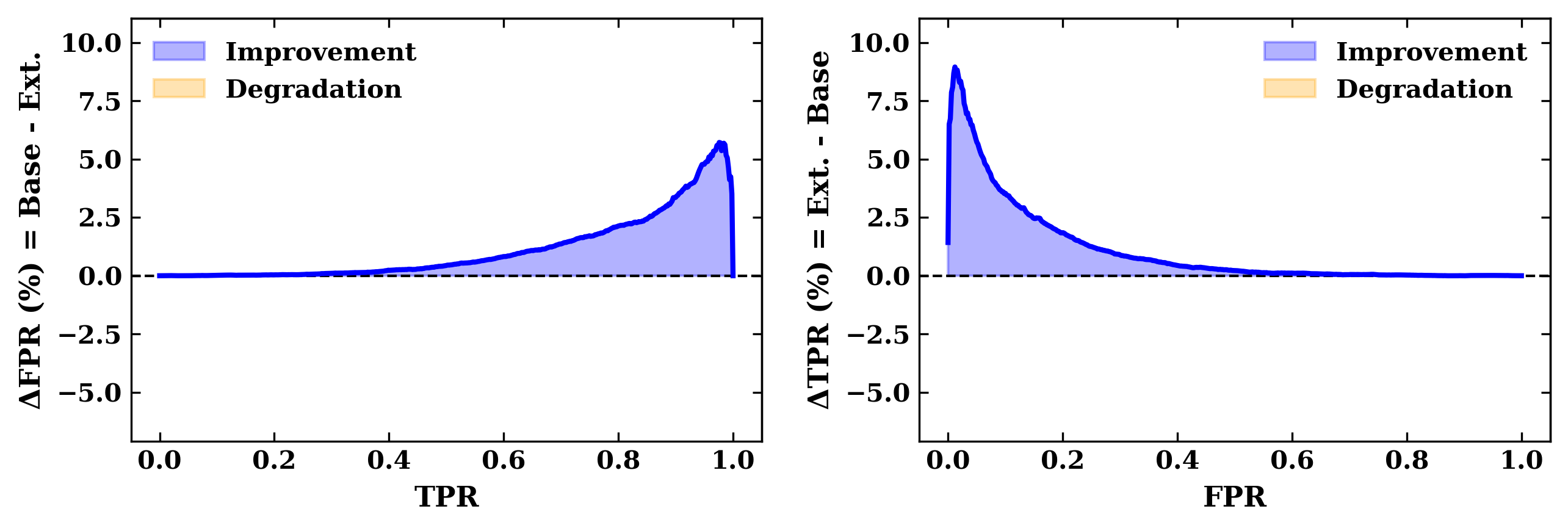}
\caption{(color online)Performance gain of the \emph{extended} BDT relative to the \emph{baseline} configuration for the electron (top) and muon (bottom) channels.
For each channel, the left panel shows the change in the background misidentification rate
$\Delta\mathrm{FPR} = \mathrm{FPR}_{\mathrm{base}} - \mathrm{FPR}_{\mathrm{ext}}$
as a function of the signal efficiency.
The right panel displays the corresponding change in signal efficiency, $\Delta\mathrm{TPR} = \mathrm{TPR}_{\mathrm{ext}} -\mathrm{TPR}_{\mathrm{base}}$, as a function of the background rate.
The shaded regions highlight the phase-space where the \emph{extended} classifier outperforms (blue) or underperforms (orange) the \emph{baseline} model.
}
\label{fig:gain_in_bdt}
\end{figure}

The TPR gain is defined as $\mathrm{TPR}_{\mathrm{ext}} - \mathrm{TPR}_{\mathrm{base}}$ again expressed in percentage such that positive values correspond to an increase in signal efficiency at fixed background. In the considered phase-space region, the \emph{extended} classifier achieves a maximum reduction in the background misidentification rate of $\Delta\mathrm{FPR}_{\max}^{e} = 5.13\%$ in the electron channel and $\Delta\mathrm{FPR}_{\max}^{\mu} = 5.71\%$ in the muon channel. The corresponding maximum gains in signal efficiency are $\Delta\mathrm{TPR}_{\max}^{e} = 8.22\%$ and
$\Delta\mathrm{TPR}_{\max}^{\mu} = 9.95\%$, respectively (see Figure~\ref{fig:gain_in_bdt}).

\subsubsection{Over-training Check}
Possible over-training of the classifier is assessed by comparing the training and independent test sample output distributions~\cite{Hocker2007TMVA_overtraining_bdt_auc_roc} evaluated using the Kolmogorov--Smirnov (KS) and $\chi^{2}$ tests for both the electron and muon channels~\cite{Kolmogorov1933_ks, Smirnov1948_ks, Massey1951_ks}.
In both channels, the KS statistics for signal and background are found to be the order of $10^{-2}$ with correspondingly small $\chi^{2}$ values indicating only minor shape differences between the training and test BDT output distributions (see table ~\ref{tab:bdt_overtraining_and_auc}). The training and test score distributions show good agreement over the full range of the classifier output, as quantified using Kolmogorov–Smirnov and $\chi^{2}$ tests (see Table~\ref{tab:auc_overtraining_summary}).
In addition, the ratios of test to training distributions remain close to unity across the full BDT score range, demonstrating uniform consistency between the samples (see Figure~\ref{fig:performance}). Taken together, these observations indicate stable classifier behaviour with no evidence of significant over-training and confirm that the discrimination power reflects genuine physical differences between signal and background.

\begin{table}[H]
\centering
\footnotesize
\renewcommand{\arraystretch}{1.25}
\begin{tabular}{|c|c|c|c|c|c|c|c|}
\hline
\textbf{Channel} & \textbf{Sample} &
$\boldsymbol{\mathrm{KS}_{\text{base}}}$ &
$\boldsymbol{\mathrm{KS}_{\text{ext}}}$ &
$\boldsymbol{\Delta\mathrm{KS}}$ &
$\boldsymbol{\chi^{2}_{\text{base}}}$ &
$\boldsymbol{\chi^{2}_{\text{ext}}}$ &
$\boldsymbol{\Delta\chi^{2}}$ \\
\hline
Electron & Signal     & 0.0134 & 0.0149 & +0.0015 & 0.2481 & 0.3284 & +0.0803 \\
& Background & 0.0102 & 0.0139 & +0.0037 & 0.1394 & 0.1988 & +0.0594 \\
\hline
Muon & Signal     & 0.0103 & 0.0110 & +0.0007 & 0.1635 & 0.2248 & +0.0613 \\
& Background & 0.0083 & 0.0110 & +0.0027 & 0.1023 & 0.1540 & +0.0517 \\
\hline

\end{tabular}
\caption{Summary of over-training diagnostics with classification performance for the \emph{baseline} and \emph{extended} BDT classifiers, where Kolmogorov–Smirnov (KS) $\chi^{2}$ tests quantify agreement between training test samples for signal background for the electrons and muons.}
\label{tab:bdt_overtraining_and_auc}
\end{table}

\section{Background Suppression Strategy and Signal Purity Enhancement}
\label{sec:Signal-to-Background}

The Drell-Yan ($Z/\gamma^{*}$) background in the invariant mass range $60 < m_{\ell\ell} < 120~\mathrm{GeV}$ poses a significant challenge to the $H \rightarrow Z\gamma$ analysis due to its kinematic properties closely resembling those of the signal. In this region, the Drell-Yan process constitutes the primary limitation to the analysis sensitivity due to its substantial overlap with the signal in both mass and kinematic observables. In the following sections, we propose a background-rejection strategy and evaluate it through a structured analysis approach in which the different stages of event selection are treated consistently and studied independently. The analysis is organised into two complementary configurations, referred to hereafter as the \emph{baseline} and the \emph{background-rejected} selection analyses. 

The \emph{baseline} configuration, corresponding to the selections described in Section~\ref{sec:event_selection}, defines the reference signal and background samples serving as the starting point for all subsequent studies. Building upon this baseline, a second \emph{background-rejected} analysis configuration is constructed by applying an additional background rejection criterion on top of the baseline selection, based on the observation that signal and background events exhibit distinctly different population patterns in the $(\theta_{Z\gamma}, P_{\mathrm{Higgs}})$ plane, as illustrated in Figure~\ref{fig:2d_plot}. The two configurations differ only in the presence of the momentum-dependent angular rejection criteria applied to background-dominated regions. The distributions in the $(\theta_{Z\gamma}, P_{\mathrm{Higgs}})$ plane, the construction of background rejection contours, the resulting signal retention, background rejection efficiency, and the improvement in signal purity using the reconstructed $m_{\ell\ell\gamma}$ distribution are discussed in the following.

\paragraph{Event Weighting}
 Signal and background samples are normalised to a common integrated luminosity of $139~\mathrm{fb}^{-1}$, corresponding to the full Run~2 dataset. For each event, a weight is assigned to convert the generated Monte Carlo sample into the expected number of events in the data. The event weight is defined as equation~\ref{equation:weights}.
\begin{equation}
    w = \frac{\sigma \times \mathcal{L}}{N_{\mathrm{gen}}}
    \label{equation:weights}
\end{equation}

where $\sigma$ denotes the production cross section of the corresponding process, $\mathcal{L}$ is the target integrated luminosity, and  $N_{\mathrm{gen}}$ is the total number of generated events, ensuring that the results correspond to physically meaningful luminosity-scaled expectations. Consequently, any change in the composition of the selected event sample originates solely from the applied selection strategy, not from differences in event normalisation.

\subsection{Construction of Background-Rejection Selections}
\label{subsec:correlation_bkg_rejection}
To identify regions where background can be effectively suppressed, the two-dimensional distributions in the $(P_{\mathrm{Higgs}},\,\theta_{Z\gamma})$ plane are examined separately for signal and background events in both the $e^{+}e^{-}\gamma$ and $\mu^{+}\mu^{-}\gamma$ final states. These distributions, shown in Figure~\ref{fig:2d_plot}, are evaluated within the phase space defined by the baseline event selection described in Section~\ref{sec:event_selection}.

\subsubsection{Key Differences in $(P_{\mathrm{Higgs}},\theta_{Z\gamma})$ plane for Signal and Background}

In both leptonic channels, signal events populate a compact and well-defined region in the $(P_{\mathrm{Higgs}}, \theta_{Z\gamma})$ plane, exhibiting a clear correlation between the Higgs boson momentum and the opening angle. In contrast, background events are strongly enhanced at large values of $P_{\mathrm{Higgs}}$ ($P_{\mathrm{Higgs}} \gtrsim 600~\mathrm{GeV}$), where they dominate over the signal contribution. The signal, by contrast, is concentrated in a more localised region characterised by moderate Higgs momenta ($P_{\mathrm{Higgs}} \lesssim 300~\mathrm{GeV}$) and central angular values $0.5 \lesssim \theta_{Z\gamma} \lesssim 1.5~\mathrm{rad}$. Furthermore, the background exhibits a forward--backward peaking behaviour with increased density near $\theta_{Z\gamma}\approx 0$ and $\theta_{Z\gamma}\approx \pi$, while the signal shows a significantly flatter and more central angular distribution. The construction of a background-rejection strategy that identifies and removes background-enriched regions while preserving signal-populated regions in the 2D distribution is presented in the subsequent section.

\subsubsection{Background fitting and rejection in the $(P_{\mathrm{Higgs}}, \theta_{Z\gamma})$ phase space} 

To identify and reject the background-dominated regions in the $(P_{\mathrm{Higgs}},\,\theta_{Z\gamma})$ plane, a bin-by-bin maximum population extraction procedure is employed. For each bin in $P_{\mathrm{Higgs}}$, the corresponding distribution of $\theta_{Z\gamma}$ is examined separately for signal and background, and the bin with the maximum population is identified. The background distribution is observed to be more diffuse, confirming that it does not exhibit the exact kinematic correlation as the signal. Given this behaviour, the strategy adopted here is to identify and selectively reject regions dominated by background while retaining events in the signal-populated region. For this reason, the extraction and subsequent fitting procedure focuses on characterising the background peak positions. The local spread around the background peak in each $P_{\mathrm{Higgs}}$ slice is labelled as $\sigma$, estimated using the Root Mean Squared(RMS) of the neighbouring bin contents, providing an empirical measure of the width of the background-dominated band.
\begin{figure}[hbt!]

    \centering
    \begin{minipage}{0.48\textwidth}
        \centering
        \includegraphics[width=\textwidth]{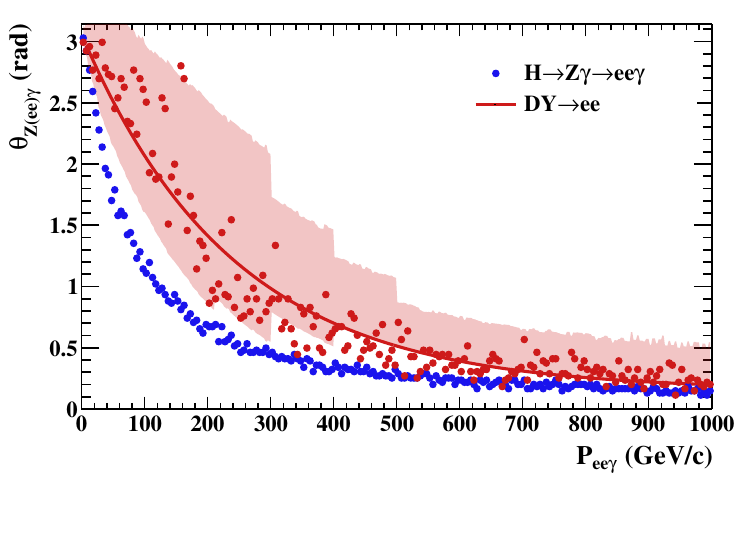}

    \end{minipage}
     \hfill
    \begin{minipage}{0.48\textwidth}
        \centering
        \includegraphics[width=\textwidth]{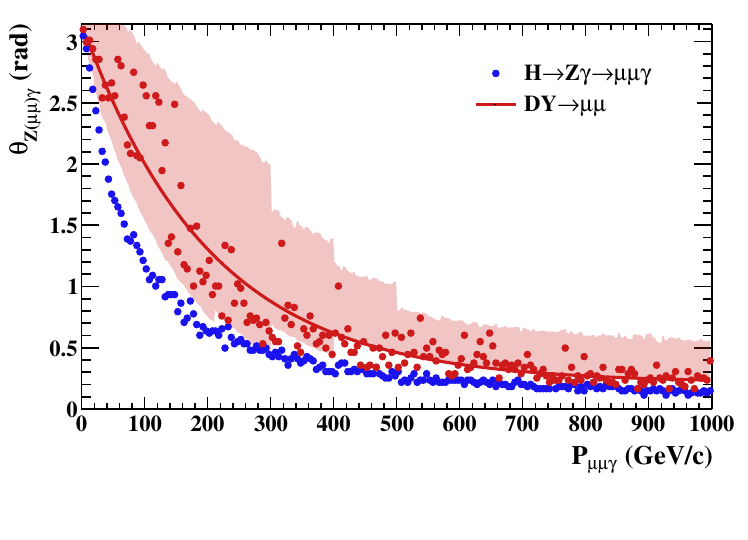}

    \end{minipage}

    \caption{(color online)The fitted background trend in the $(\theta_{Z\gamma}, P_{\mathrm{Higgs}})$ plane is shown along with the corresponding $n\sigma$ rejection bands (\emph{red}). These bands define background-dominated regions of the phase space and form the basis of the background rejection strategy employed in this analysis.  
        }
    \label{fig:2d_plot_and_extraction_of_signal}
\end{figure}
The sequence of extracted background peak positions as a function of $P_{\mathrm{Higgs}}$, shown in Figure~\ref{fig:2d_plot_and_extraction_of_signal}, exhibits a characteristic behaviour consisting of a steep decrease at low momenta followed by a gradual flattening at higher $P_{\mathrm{Higgs}}$.
This trend is consistent across both the electron and muon channels. The extracted background peak positions are subsequently fitted using an exponential function in the equation~\ref{equation:f(x)}.

\begin{equation}
    f(x) = A\,e^{-Bx} + C
    \label{equation:f(x)}
\end{equation}

which models a rapidly falling component that saturates towards a constant value at large $P_{\mathrm{Higgs}}$, yielding $\chi^{2}/\mathrm{ndf} < 1$~\cite{PDG2024Statistics}. Around the fitted background trend, asymmetric rejection bands are constructed using the locally determined standard deviations $\sigma$'s. The upper and lower boundaries, expressed in units of $\sigma$, are chosen independently in each $P_{\mathrm{Higgs}}$ interval based on the observed signal and background populations. This procedure is purely empirical and does not rely on extracting a signal correlation or any fixed analytical formula. Instead, the boundaries are adjusted to maximise background rejection while retaining a high fraction of signal events, ensuring an optimal balance between background suppression and signal retention. Having identified the background-dominated regions in the $(P_{\mathrm{Higgs}},\,\theta_{Z\gamma})$ plane through fitting, the next step is to quantify the performance of this rejection strategy in terms of signal retention and background rejection, as presented in Section~\ref{subsec:sig_bkg_eff}.

\subsection{Signal Retention and Background Rejection Efficiency}
\label{subsec:sig_bkg_eff}

Following the identification of background-dominated regions through fitting described in Section~\ref{subsec:correlation_bkg_rejection}, the performance of the background rejection strategy is quantified in terms of the signal retention efficiency and the background rejection efficiency described in equation~\ref{eq:eff_rej}, evaluated as functions of the rejection bands defined in the $(\theta_{Z\gamma}, P_{\mathrm{Higgs}})$ plane. The resulting efficiencies for the chosen $n\sigma$ ranges are defined as and summarised in Table~\ref{tab:sig_back_eff}.

\begin{table}[H]
\centering
\footnotesize
\renewcommand{\arraystretch}{1.25}

% ------------------- ELECTRON TABLE -------------------
\begin{tabular}{c c c c}
\hline
\textbf{Cut Set} & \textbf{$P_{\mathrm{H}}$ Range (GeV)} 
& \textbf{Signal Efficiency (\%)} 
& \textbf{Background Rejection (\%)} \\
\hline
1 & 0--210    & 76.3205 & 67.1618 \\
2 & 210--300  & 67.3106 & 67.3296 \\
3 & 300--400  & 81.0761 & 45.1003 \\
4 & 400--500  & 91.0066 & 28.5656 \\
5 & 500--1000 & 79.646 & 34.8253 \\
\hline
\end{tabular}

\vspace{0.6cm}

% ------------------- MUON TABLE -------------------
\begin{tabular}{c c c c}

\hline
\textbf{Cut Set} & \textbf{$P_{\mathrm{H}}$ Range (GeV)} 
& \textbf{Signal Efficiency (\%)} 
& \textbf{Background Rejection (\%)} \\
\hline
1 & 0--210    & 67.2393 & 71.3325 \\
2 & 210--300  & 48.553 & 74.74 \\
3 & 300--400  & 82.8669 & 40.4547 \\
4 & 400--500  & 78.4273 & 40.4681 \\
5 & 500--1000 & 73.4273 & 41.1794 \\
\hline

\end{tabular}

\caption{
Asymmetric $\pm n\sigma$ selection applied to the 
$\theta_{Z\gamma}$-$P_{\mathrm{Higgs}}$ plane for the 
$e^{+}e^{-}\gamma$ and $\mu^{+}\mu^{-}\gamma$ channels.  
For each $P_{\mathrm{H}}$ interval  
$[0,210]$, $[210,300]$, $[300,400]$, $[400,500]$ and $[500,1000]$~GeV,  
the corresponding asymmetric Gaussian windows  
($+3\sigma,-1.5\sigma$), ($+3\sigma,-1.2\sigma$),  
($+2\sigma,-0.5\sigma$), ($+1.5\sigma,-0\sigma$), and ($+1\sigma,-0\sigma$)  
are applied respectively.  
The tables summarise the resulting signal efficiency and Drell-Yan background rejection efficiencies for both electron and muon final states.}
\label{tab:sig_back_eff}
\end{table}

\begin{equation}
\epsilon_{S}(n\sigma) = 1 - \frac{S_{\mathrm{in}}(n\sigma)}{S_{\mathrm{tot}}},
\qquad
\epsilon_{B}(n\sigma) = \frac{B_{\mathrm{in}}(n\sigma)}{B_{\mathrm{tot}}}
\label{eq:eff_rej}
\end{equation}

Here, $S_{\mathrm{in}}(n\sigma)$ and $B_{\mathrm{in}}(n\sigma)$ denote the signal and background yields obtained within the background-dominated region defined by the $n\sigma$ rejection criterion. These yields are computed from the reconstructed $m_{\ell\ell\gamma}$ distributions after applying appropriate event weighting based on the production cross sections and the integrated luminosity. The corresponding total signal and background yields, $S_{\mathrm{tot}}$ and $B_{\mathrm{tot}}$, are defined as the integrals of the same weighted $m_{\ell\ell\gamma}$ distributions over the full baseline phase-space region under consideration.

The results presented in Table~\ref{tab:sig_back_eff} demonstrate that the background rejection strategy achieves substantial background suppression across all $P_{\mathrm{Higgs}}$ intervals while maintaining high signal retention. These results confirm the effectiveness of the momentum-dependent background rejection strategy, supporting the observation that background events do not exhibit the same exact kinematic correlation as signal events. As changes in the integrated luminosity therefore lead only to a global rescaling of the event yields, the efficiencies remain unchanged as long as the same rejection criteria are applied, which is particularly advantageous for future high-statistics datasets, as the same background rejection performance is expected to persist during the full Run~3 and HL-LHC data-taking periods~\cite{Apollinari2017}.

\subsection{Signal Purity Improvement}
\label{subsec:SB_ratio_improvement}

The signal retention efficiency and background rejection efficiency presented in Section~\ref{subsec:sig_bkg_eff} quantify the performance of the background rejection strategy as a function of $n\sigma$ contour. The ultimate impact on analysis sensitivity is assessed by evaluating the signal purity, defined as $(S+B)/B$, in the reconstructed Higgs boson mass spectrum. The effect of excluding background-dominated regions in the $(P_{\mathrm{Higgs}}, \theta_{Z\gamma})$ plane is assessed by comparing the signal purity obtained before and after applying the asymmetric $P_{\mathrm{Higgs}}$ dependent $n\sigma$ rejection bands described in Section~\ref{subsec:correlation_bkg_rejection}. To ensure that any observed improvement in purity originates solely from the background rejection strategy, event weighting and mass reconstruction procedures are kept unchanged.

\subsubsection{$(S+B)/B$ Extraction Methodology}
\begin{figure}[H]
\centering
\begin{tikzpicture}[scale=1.05]

% axes
\draw[->] (0,0) -- (8,0) node[right] {$m_{\ell\ell\gamma}$};
\draw[->] (0,0) -- (0,5.0) node[above] {Events};

% background curve
\draw[line width=1pt, blue!60]
plot[smooth] coordinates {(0.3,2.8) (1,2.4) (2,1.9) (3,1.5) (4,1.25) (5,1.1) (6,1.05) (7,1.0)};
\node[blue!70] at (6.2,2.2) {B(x)};

% signal peak
\draw[line width=1pt, red!70]
plot[smooth] coordinates {(2.7,1.5) (2.9,1.6) (3.1,2.6) (3.35,3.6)
(3.55,2.55) (3.8,1.5) (4.0,1.2)};
\node[red!70] at (2.5,3.2) {S(x)};

% combined curve (merges back with background)
\draw[line width=1pt, black]
plot[smooth] coordinates {(0.3,2.8) (1,2.45) (2,2.1) (2.8,2.2)
(3.0,2.9) (3.2,3.8) (3.4,4.1)
(3.6,3.0) (3.8,2.2) (4.0,1.4)
(5,1.1) (6,1.05) (7,1.0)};
\node[black] at (4.2,3.2) {F(x)};

\end{tikzpicture}

\caption{(color online)Schematic illustration of the fit model components: the background polynomial $B(x)$ (blue), the signal Gaussian $S(x)$ (red) and their sum $F(x)$ (black) representing the total fitted distribution. The signal purity ratio $(S+B)/B$ is computed bin-by-bin by evaluating $F(x)$ and $B(x)$ at the center of each mass bin, and the maximum ratio is reported.}
\label{fig:schematic_fit}
\end{figure}
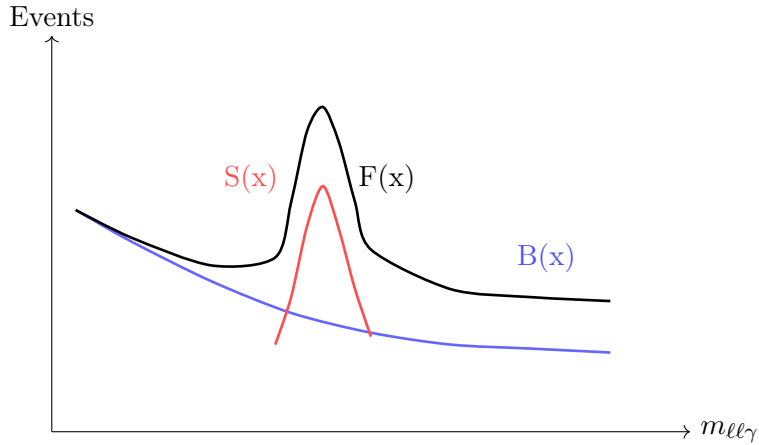

The signal purity is quantified using the ratio $(S+B)/B$ extracted from a bin-by-bin calculation in the reconstructed $m_{\ell\ell\gamma}$ distribution. This provides a direct measure of how much the observed event yield exceeds the background expectation in each mass bin, serving as a quantitative indicator of signal presence or of improved purity. The maximum value of this ratio across all bins in the Higgs boson mass region is used as the representative measure of signal purity. The ratio calculation employs two components fitted to the $m_{\ell\ell\gamma}$ spectrum: a smooth polynomial function to model the background continuum and a Gaussian function to describe the signal peak. The combined model is defined as a set of equations.
\begin{equation}
B(x) = p_{0} + p_{1}x + p_{2}x^{2} + p_{3}x^{3}
\label{eq:background}
\end{equation}

\begin{equation}
S(x) = A \exp\!\left[-\frac{1}{2}
\left(\frac{x - \mu}{\sigma}\right)^{2}\right]
\label{eq:signal}
\end{equation}

\begin{equation}
F(x) = B(x) + S(x)
\label{eq:total}
\end{equation}
As shown in Eq.~(\ref{eq:background}), the background is modeled by a third-order polynomial. 
The signal contribution, given in Eq.~(\ref{eq:signal}), is described by a Gaussian function. 
The total model in Eq.~(\ref{eq:total}) represents the sum of both components. The coefficients $p_{1}$, $p_{2}$, and $p_{3}$ control the linear, quadratic, and cubic components, respectively, allowing the model to capture the smooth shape and curvature of the background continuum across the fitted range. The parameter $A$ denotes the signal amplitude, $\mu$ corresponds to the mean of the Gaussian and represents the reconstructed Higgs boson mass, and $\sigma$ characterises the width of the signal peak, reflecting the experimental mass resolution. An illustration of this fitting procedure is shown in Figure~\ref{fig:schematic_fit}.

The signal purity ratio is computed on a bin-by-bin basis by evaluating both the combined model $F(x)$ and the background model $B(x)$ at the center of each mass bin and taking their ratio and described in equation~\ref{eq:signal_purity}.
\begin{equation}
    \left( \frac{S+B}{B} \right)_{\mathrm{bin},i}
= \frac{F(x_i)}{B(x_i)}
\label{eq:signal_purity}
\end{equation}

where $x_i$ represents the center of the $i$-th bin in the $m_{\ell\ell\gamma}$ distribution, $F(x_i)$ is the expected total yield (signal + background) evaluated at that bin center and $B(x_i)$ is the expected background only yield. The maximum value of this ratio across all bins in the Higgs boson mass region is then identified and used as the representative measure of signal purity described in equation~\ref{eq:signal_purity_max}.
\begin{equation}
\left( \frac{S+B}{B} \right)_{\mathrm{max}}
= \max \left\{ \frac{F(x_i)}{B(x_i)} \right\}
\label{eq:signal_purity_max}
\end{equation}

This directly probes the local signal-to-background enhancement at each mass point, with the maximum ratio typically occurring near the Higgs mass peak, where the signal contribution is most significant relative to the smooth background. A value of $(S+B)/B = 1$ in a given bin indicates no signal excess, while values greater than 1 indicate signal presence, with higher values corresponding to better signal purity. The improvement in the maximum ratio after applying the background rejection criteria directly reflects the effectiveness of the selection strategy in enhancing signal purity.

\subsubsection{Results and Discussion}

\begin{figure}[H]
\centering

%====================== FIRST ROW — BEFORE ======================%
\begin{minipage}{0.48\textwidth}
\centering
\includegraphics[width=\textwidth]{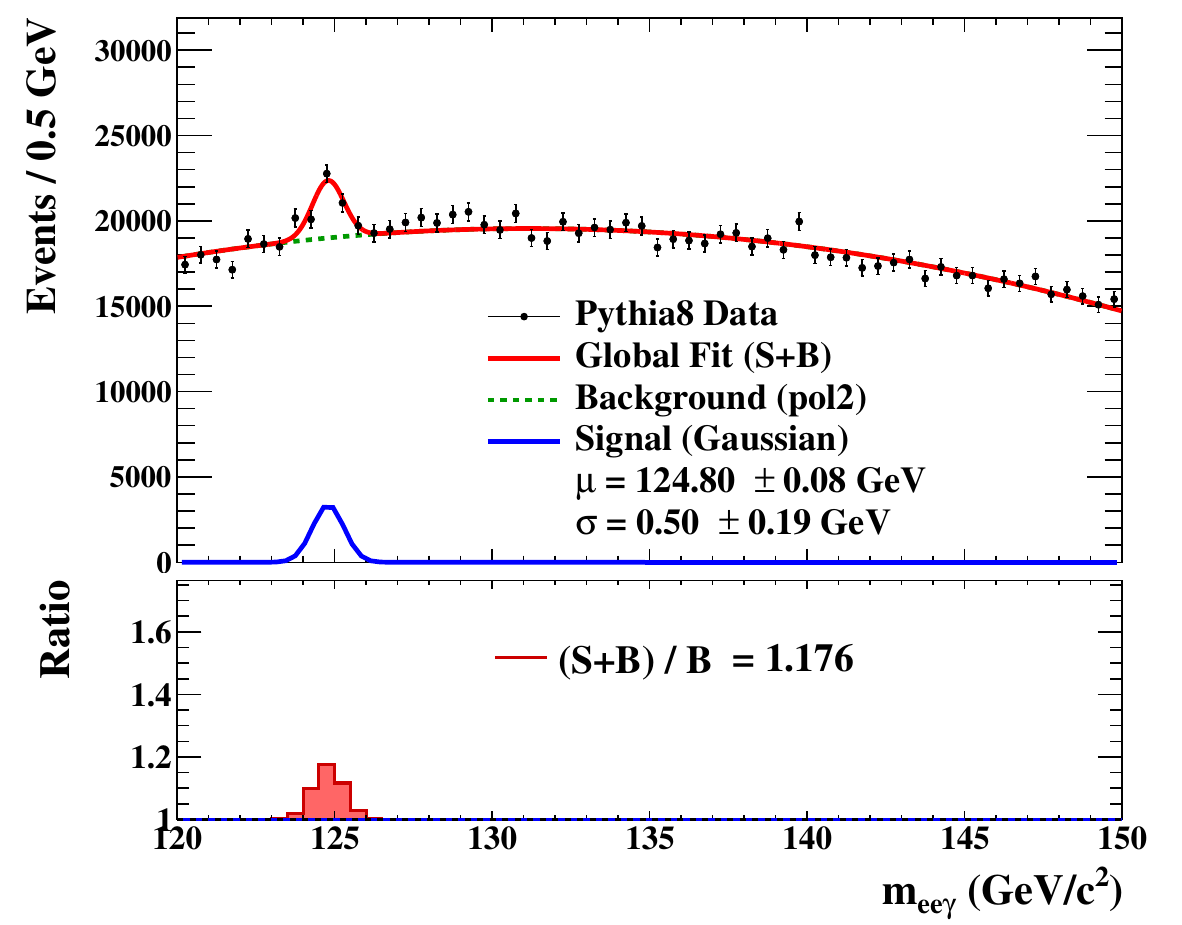}

\end{minipage}
\hfill
\begin{minipage}{0.48\textwidth}
\centering
\includegraphics[width=\textwidth]{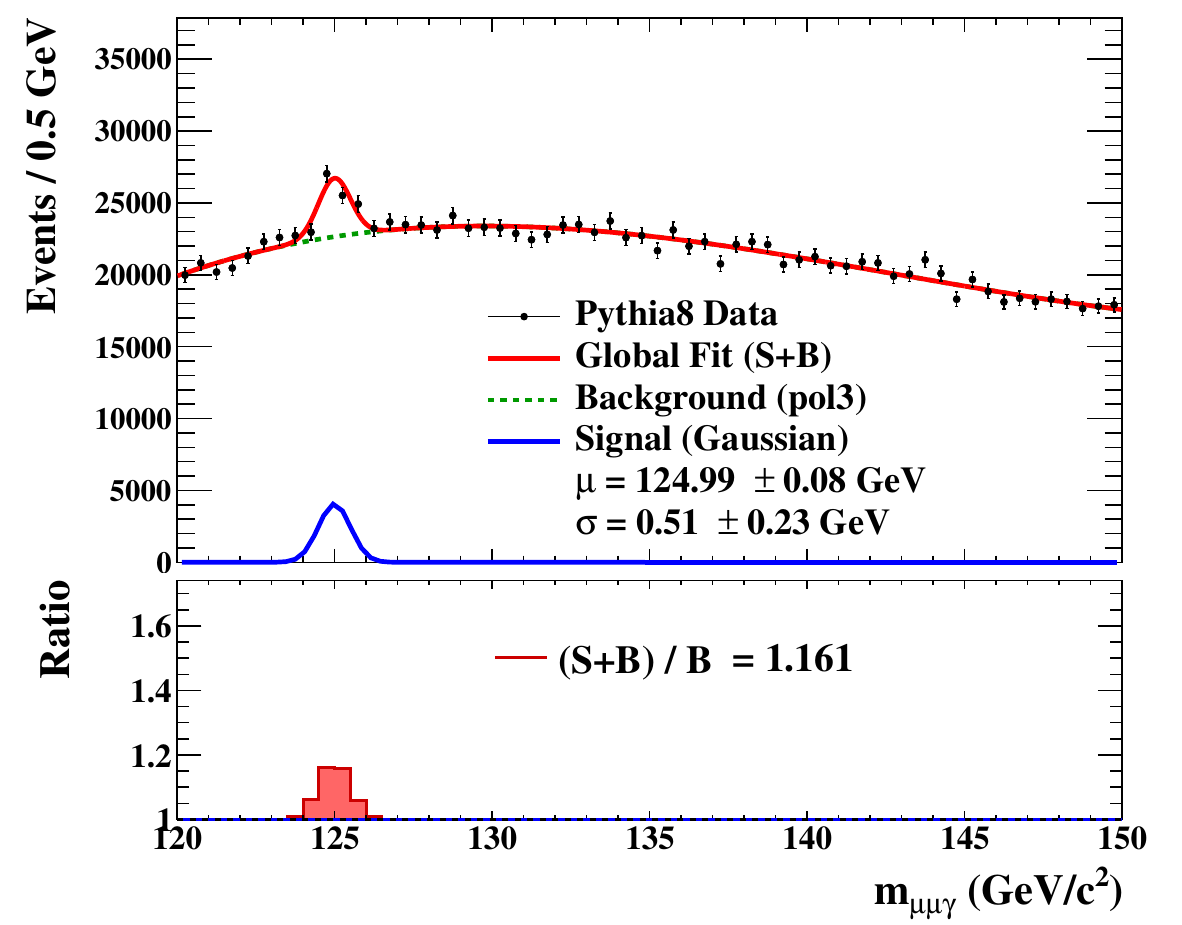}

\end{minipage}

\vspace{0.5cm}

%====================== SECOND ROW — AFTER ======================%
\begin{minipage}{0.48\textwidth}
\centering
\includegraphics[width=\textwidth]{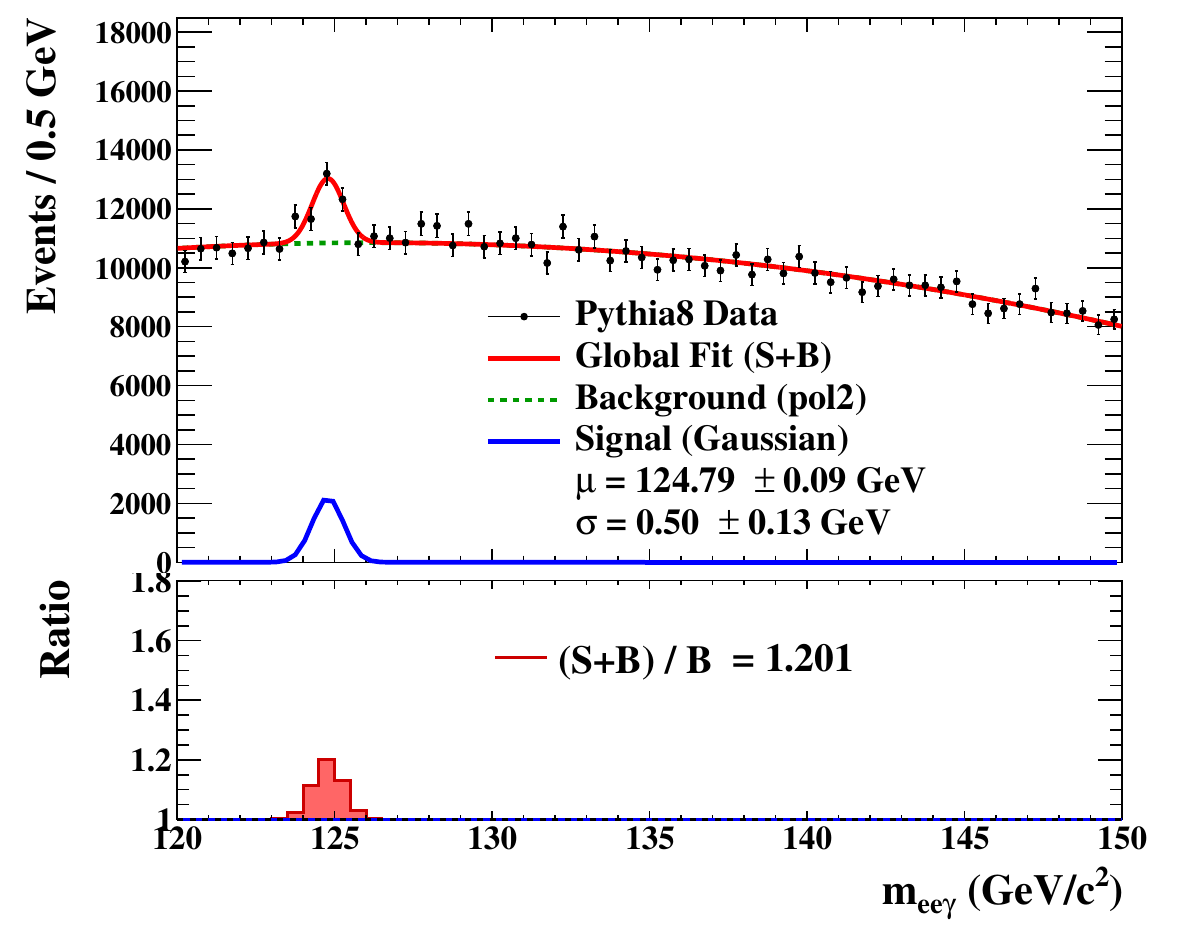}

\end{minipage}
\hfill
\begin{minipage}{0.48\textwidth}
\centering
\includegraphics[width=\textwidth]{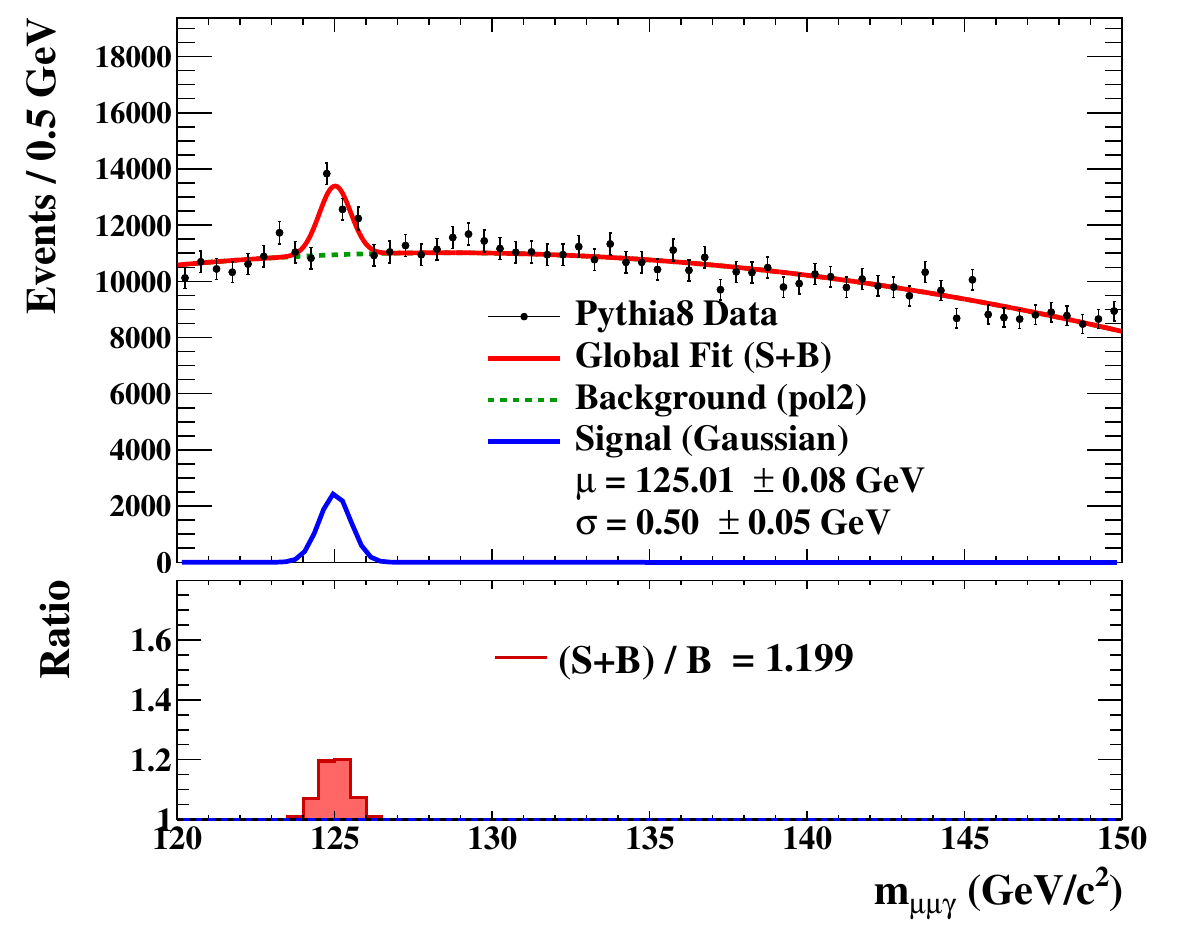}

\end{minipage}

%====================== MAIN CAPTION ======================%
\caption{(color online)Signal purity ratio $(S+B)/B$ in the Higgs mass region for the electron(\emph{left coloumn}) and muon {\emph{right coloumn}} channels, shown separately before (top row) and after (bottom row) the application of the momentum--dependent background rejection strategy derived from fitting the background-dominated regions in the \(P_{\mathrm{Higgs}}\)--\(\theta_{Z\gamma}\) plane.}
\label{fig:sb_ratio_4panel}
\end{figure}

The comparison of the maximum signal purity ratios before and after the application of the background rejection strategy is presented in Figure~\ref{fig:sb_ratio_4panel} for both the electron and muon channels. 

A clear improvement in the sample purity is observed in both channels. The asymmetric rejection of $Z/\gamma^*$ events in each $P_{\mathrm{Higgs}}$ bin based on the fitted background-dominated regions leads to a substantial suppression of the background under the Higgs mass peak. At the same time, the associated signal loss remains small due to the momentum-dependent optimisation of the rejection bands. Quantitatively, the maximum signal purity ratio changes from $(S+B)/B = 1.176$ to $(S+B)/B = 1.201$ in the electron channel and from $(S+B)/B = 1.161$ to $(S+B)/B = 1.200$ in the muon channel after applying the $P_{\mathrm{Higgs}}$--$\theta_{Z\gamma}$ based rejection, corresponds to an enhancement of approximately $2.1\%$ in the electron channel and $3.4\%$ in the muon channel, representing a meaningful improvement in signal purity. These results further indicate that the background does not follow the kinematic correlation observed in the signal in the $(P_{\mathrm{Higgs}}, \theta_{Z\gamma})$ plane, enabling selective rejection of background-dominated regions via a momentum-dependent fitting strategy. Combined with the efficiency studies, this approach provides a robust improvement in the sensitivity of the $H \to Z\gamma$ search in the presence of dominant Drell-Yan backgrounds, without significantly distorting the signal statistics.

\subsubsection{Interpretation in the Context of Small Signal Rates}

\begin{figure}[H]
\centering
\includegraphics[width=1\textwidth]{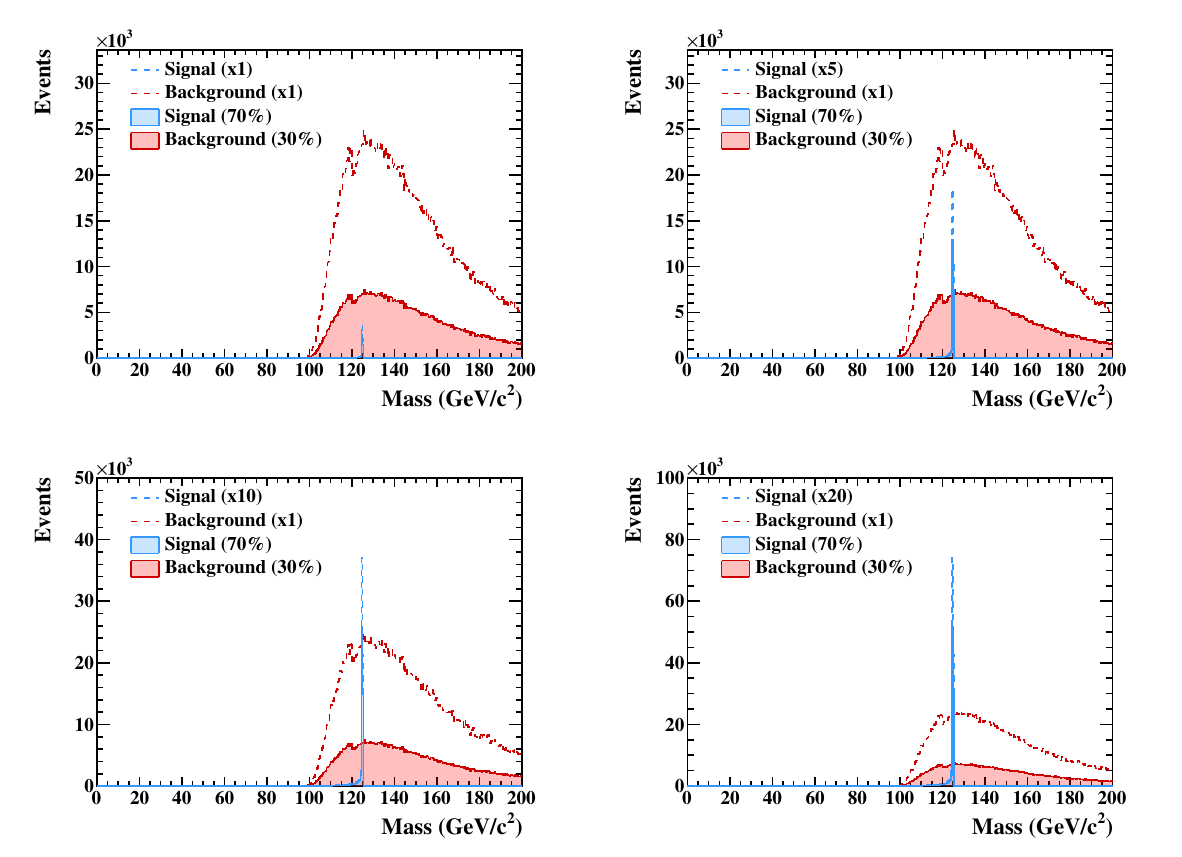}
\caption{(color online)Illustration of the effect of the $P_{\mathrm{Higgs}}$--$\theta_{Z\gamma}$ 
background rejection on the $m_{\ell\ell\gamma}$ spectrum for different signal 
normalisations.  
The top-left panel (S$\times$1, B$\times$1) corresponds to the physical 
configuration with the SM $H\to Z\gamma$ cross section and the nominal 
Drell-Yan background. In this case, the 30\% signal loss and 70\% background 
reductions are not visually prominent because the signal is much smaller 
than the background.
In the other panels (S$\times$5, S$\times$10, S$\times$20, with B$\times$1), the signal yield is artificially scaled by constant factors for illustration, while the same rejection criteria and background normalisation are maintained. The expected behaviour becomes visible: the signal peak is moderately reduced, whereas the Drell-Yan continuum is strongly suppressed.}
\label{fig:scaled_mass_panels}
\end{figure}

The impact of the background rejection on the $H\to Z\gamma$ signal is not visually striking in the nominal configuration due to the small Standard Model signal rate relative to the Drell-Yan background. To illustrate the effectiveness of the rejection strategy more clearly, Figure~\ref{fig:scaled_mass_panels} shows the $m_{\ell\ell\gamma}$ distributions for different signal normalisations. 
The top-left panel (S$\times$1, B$\times$1) corresponds to the physical normalisation, where the Higgs signal cross section is much smaller than the Drell-Yan background. Here, the quoted background and signal reductions correspond to illustrative efficiency choices made for visualisation and are not visually prominent on the large Drell-–Yan spectrum. To make the effect of the selection easier to see, the remaining panels (S$\times$5, S$\times$10, S$\times$20, with B$\times$1) are obtained by artificially scaling the signal yield by constant factors while keeping the background fixed. No additional tuning of the rejection criteria is performed, and the same efficiencies are applied in all cases; only the signal normalisation is changed for visualisation purposes. As the signal is scaled up, the expected pattern becomes clear: the peak is reduced by the 30\% signal loss, while the 70\% background rejection strongly suppresses the Drell-Yan continuum, demonstrates that the slight change in $(S+B)/B$ observed in the physical (S$\times$1, B$\times$1) scenario is a consequence of the small SM $H \to Z\gamma$ rate, not a limitation of the rejection strategy itself.

\section{Systematic Uncertainties and Experimental Considerations}
\label{sec:systematics}

In this work, we focus on generator-level (MC-level) kinematic properties to assess the potential improvement in signal ($H \to Z\gamma \to \ell^+\ell^-\gamma$) and background discrimination. The analysis is primarily performed using a boosted decision tree (BDT) classifier, which enables a clear evaluation of the intrinsic discriminating power of the chosen observables. In addition to the multivariate approach, we also investigate the use of correlated kinematic observables to enhance the $(S+B)/B$ ratio. It is possible to identify regions of phase space where the signal contribution is relatively enhanced compared to the background  with correlations among key variables of the dilepton+photon system. A complete experimental analysis, however, requires the inclusion of several important systematic effects. One of the primary sources of uncertainty arises from detector effects, particularly the energy scale and resolution of photons and leptons, which directly impact the kinematic observables used in this analysis. In realistic detector environments, such effects lead to a smearing of distributions, potentially reducing the separation between signal and background. However, since the present study use global kinematic features and correlations of the dilepton+photon system, rather than relying solely on sharp resonant structures, the overall discriminating power is expected to be relatively robust against moderate detector smearing~\cite{CMS:2008xjf,CMS:2015myp}. Another important consideration is the impact of higher-order QCD corrections. Next-to-leading order (NLO) and beyond can modify both the normalization and shapes of kinematic distributions for signal and background processes. These corrections can affect jet activity, recoil, and transverse momentum spectra. Nevertheless, the observables used in this study are constructed from the final-state dilepton+photon system, which is expected to retain its overall kinematic characteristics under such corrections. Therefore, while higher-order effects may influence absolute rates, their impact on the relative discrimination power is expected to be subdominant~\cite{Alwall:2014hca,Frixione:2002ik}. In addition, Higgs boson production occurs via multiple channels at the LHC, including gluon fusion, vector boson fusion, and associated production. The present study does not explicitly separate these production modes, as it primarily targets the decay topology of the $H \to Z\gamma$ system. The inclusion of different production mechanisms would mainly alter the overall event yields and potentially introduce variations in event topology, but is not expected to significantly change the discriminating features of the decay-level observables used in the classifier~\cite{LHCHiggsXS}. A further important background in experimental analyses arises from jets misidentified as photons. Such ``fake photon'' contributions are typically non-negligible and depend strongly on detector performance, photon identification criteria, and data-driven background estimation techniques. Since this study is performed at generator level, only prompt photons are considered. Consequently, the impact of jet-induced fake photons is not included here and should be accounted for in a full detector-level implementation~\cite{CMS:2015myp,ATLAS:2014fak}. Overall, while the present analysis provides a proof-of-principle demonstration of improved discrimination using machine learning and kinematic observables, the inclusion of detector effects, higher-order corrections, additional production channels, and instrumental backgrounds will be essential for a realistic experimental application. These aspects are left for future work.

%======================
\section{Summary}
\label{sec:summary}
The study demonstrates the effectiveness of kinematic correlations in
$H \rightarrow Z\gamma \rightarrow \ell^+\ell^-\gamma$ signal production at $\sqrt{s}=13~\text{TeV}$. It identifies regions of the $(\theta_{Z\gamma}, P_{\mathrm{Higgs}})$ phase space where the dominant $Z/\gamma^{*}$ background can be efficiently suppressed relative to the signal. XGBoost-based BDT classifiers have been trained to discriminate the $H\to Z\gamma$ signal from the background $Z\gamma$, comparing a set of baseline characteristics of 9 kinematic variables with an enhanced set of 12 variables including the new $log(\theta_{Z\gamma} \times P_{\mathrm{Higgs}})$ observable. The enhanced feature set achieved AUC improvements of 0.0108 and 0.0110 in the electron and the muon channel, respectively, with $log(\theta_{Z\gamma} \times P_{\mathrm{Higgs}})$ identified as the most crucial discriminating variable. Over-training diagnostics confirmed an excellent train-test agreement, with KS statistic test values of $<10^{-2}$ for all models. The analysis introduces a momentum-dependent angular selection that leverages the distinct kinematic behaviour of signal and background events in the $(P_{\mathrm{Higgs}}, \theta_{Z\gamma})$ plane. For each $P_{\mathrm{H}}$ interval using an exponential form, asymmetric $\pm n\sigma$ windows are extracted to minimise signal loss while rejecting high Drell-Yan events. Across the five kinematic intervals, this method removes $\sim$70\% of the Drell-Yan contamination while preserving $\sim$70\% of the signal. The impact of the background rejection strategy on signal purity was quantified using the $(S+B)/B$ ratio extracted from bin-by-bin calculations in the reconstructed $m_{\ell\ell\gamma}$ distribution. The maximum ratio across all bins in the Higgs boson mass region was evaluated before and after applying the rejection based on $P_{\mathrm{Higgs}}$--$\theta_{Z\gamma}$. In the electron channel, the maximum $(S+B)/B$ ratio improves from 1.176 to 1.201, corresponding to a 2.1\% increase in signal purity. In the muon channel, the ratio increases from 1.161 to 1.200, representing a 3.4\% improvement. These improvements demonstrate the effectiveness of the physics-motivated background-rejection strategy in improving signal purity while maintaining signal efficiency. Because the rate of the Standard Model $H\to Z\gamma$ is extremely small, the resulting $(S+B)/B$ improvement is numerically modest in the physical (S$\times$1, B$\times$1) configuration. To illustrate the intrinsic effectiveness of the selection, additional plots are produced by scaling the signal yield by constant factors, without implying any physical result.
These figures demonstrate that the same selection would visibly enhance the signal region in scenarios where the signal is not overwhelmed by background, confirming that the limited $(S+B)/B$ improvement arises from the small Standard Model $H \to Z\gamma$ rate rather than from a weakness of the kinematic method. The methodology remains transparent, easy to validate, and robust against model assumptions. Therefore, such kinematic correlations can be used safely as discriminators in future analyses, extending beyond the $H \to Z\gamma$ channel to other resonance searches.

%======================
\section*{Acknowledgments}

The authors acknowledge valuable discussions with members of the HEP group at IIT Mandi, whose insights and feedback have significantly improved this work. The PhD fellowship support from the Department of Science and Technology (DST), Government of India, through the DST–INSPIRE program is gratefully acknowledged by the corresponding author. The computing resources provided by the Indian Institute of Technology Mandi are also acknowledged.

\bibliographystyle{JHEP}
\bibliography{biblio.bib}

\end{document}